\documentclass[a4paper,11pt]{article}
\pdfoutput=1 

\usepackage{jheppub} 

\usepackage[T1]{fontenc} 
\usepackage{subcaption}

\title{\boldmath Impact of Rare Decays $t \to \ell' \nu b \ell\ell$ and $t \to
q q' b \ell\ell$ on Searches for Top-Associated Physics}

\author{P.\ Onyisi,}
\author{A.\ Webb}

\affiliation{Department of Physics, University of Texas at Austin\\Austin,
  TX 78712, United States }

\emailAdd{ponyisi@utexas.edu}
\emailAdd{awebb@cern.ch}

\abstract{Searches for top quark-associated physics such as $t\bar t W$ or
  $t\bar t H$ in final states with multiple leptons require a careful
  accounting of expected backgrounds due to the lack of reconstructible
  resonances.  We demonstrate that the rare top quark
  decays $t \to \ell' \nu b \ell\ell$ and $t \to q q' b \ell\ell$, when a
  soft lepton is not detected, can contribute a non-negligible background to
  such searches. Simulations in the LHC experiments typically do not account
  for such decays and as such backgrounds to such searches may be underestimated.}

\begin{document} 
\maketitle
\flushbottom

\section{Introduction}
\label{sec:intro}

The achievement of 13 TeV center of mass energy and design collision
luminosity at the Large Hadron Collider (LHC) will enable a rich program of
studies of the production of electroweakly-interacting particles produced in association with
the top quark.  The processes $pp \to t\bar t Z$ and $t\bar t W$ are established \cite{Aad:2015eua,Aaboud:2016xve,Khachatryan:2015sha}, while
the production of $t\bar t H$ remains to be clearly
demonstrated \cite{Khachatryan:2014qaa,Aad:2016zqi,Aad:2015iha,Khachatryan:2016vau}. These processes will provide critical information on the
coupling of the top quark to electroweak gauge bosons and the mechanism of top quark
mass generation.

One of the most powerful ways of accessing $t\bar t W$ and $t\bar t H$
production is through final states with multiple leptons, especially three
leptons or two leptons of the same sign.  These lepton requirements are
sufficient to suppress contamination from top pair production, where at most
two same-sign leptons are expected; $t\bar t$ will typically contribute a
background only through lepton charge misreconstruction, additional leptons
from heavy hadron decay, or $t\bar t\gamma$ production with subsequent $\gamma
\to e^+ e^-$ conversion in detector material.  However, since these final
states have multiple neutrinos, separation of different processes is complex
and simulation of rare processes that mimic the signature can become very important.

The potential relevance of backgrounds with ``lost'' leptons to multilepton searches was pointed out
in the context of the search for $H \to WW^* \to \ell\nu\ell'\nu$ produced in
gluon-gluon fusion \cite{Gray:2011us}.  The production of $W\gamma^* \to \ell\nu\ell'(\ell')$,
where one of the daughters of the virtual photon fails to be reconstructed,
can mimic the dilepton signature of Higgs boson production.  Such a situation is
especially common when $m(\gamma^*)$ is low and the flight directions of the
daughter leptons are aligned with the momentum of the virtual photon in the lab frame: one
lepton is then often red-shifted to very low momentum in the lab frame and may
not pass analysis momentum cuts and not contribute enough
isolation energy to cause its pair partner lepton to appear non-isolated.  This
phenomenon of ``asymmetric internal conversion'' was found to contribute a
non-trivial background to the $H \to WW^*$ search.

In this paper, we demonstrate that a generally-neglected process, $pp \to
t\bar t$, where one top quark undergoes the rare decay $t \to \ell' \nu b
\gamma^*$ or $t \to q q' b \gamma^*$ with subsequent decay $\gamma^* \to
\ell\ell$, can induce a significant background to multilepton searches for top
quark-associated phenomena via asymmetric internal conversions.  In light of
persistent excesses over the Standard Model prediction in $t\bar t W$ and $t
\bar t H$ multilepton searches, potential additional backgrounds are important
to identify.

Since generally-used QED showering algorithms do
not include virtual photon splitting to leptons, this contribution needs to be
explicitly included in event generators as a matrix element
calculation. Recently, QED parton showers supporting $\gamma \to \ell\ell$
splitting kernels have been made public (such as in newer revisions of
\textsc{Pythia} 8 \cite{Pythia8} and the C++ version of \textsc{PHOTOS} \cite{Davidson:2010ew}). Prescriptions for
the proper matching of matrix element and parton shower event generation for
this process remain to be developed and will be critical for future precision
understanding of these rare decays, but are outside the scope of this paper.

The paper is organized as follows.  In Section \ref{sec:kinematics} the
generic behavior  expected of AIC as a function of
the momentum and mass of the virtual photon is discussed. In Section
\ref{sec:decay} the decay width for the internal conversion decay of a
top quark is determined and compared with previous predictions. In Section
\ref{sec:simulation} a realistic detector parametrization is used to study how AIC events will be reconstructed and the
resulting yields in a trilepton analysis are compared to those of the $t\bar t W$ and $t \bar t H$
production processes.

\section{\label{sec:kinematics}Asymmetric Internal Conversion Kinematics}
Asymmetric internal conversion (AIC) events $\gamma^* \to \ell\ell$ will leave
different observable signatures depending on on $m(\gamma^*)$,
$p_T(\gamma^*)$, and the angle $\theta$ between the $\ell^-$ flight direction
and the $\gamma^*$ flight direction as measured in the virtual photon rest
frame.

To gain some insight into the kinematics of asymmetric internal conversion
events, we run toy Monte Carlo of $\gamma^* \to \mu^+\mu^-$ decays for various values of $m(\gamma^*)$ and
$p(\gamma^*)$, mapping the fraction of events that pass two types of
selections:
\begin{enumerate}
\item A two-lepton selection, where both leptons are required to have momentum
  above 10 GeV, and to be separated by more than 0.3 radians (isolation);
\item A one-lepton selection, where exactly one lepton must have momentum
  above 10 GeV, and where, if the two leptons lie within 0.3 radians, the
  second is required to satisfy $p(\mathrm{sublead})/p(\mathrm{lead}) < 0.1$ (isolation).
  \end{enumerate}
In the toy MC, total momentum is used instead of transverse momentum
$p_T$ to simplify the problem; they are equivalent for $\eta \approx 0$. For each value of $m(\gamma^*)$ and $p(\gamma^*)$ tested, many potential decay angles $\cos \theta$ are chosen in the rest frame of the virtual photon, and then the system is boosted into the lab frame.  The
virtual photons are assumed to be completely transversely polarized and therefore
are generated according to the probability density
\[ P(\cos\theta) = \frac{3}{8}\cdot\frac{2 - \beta^2 + \beta^2\cos^2 \theta}{1 +
  \frac{1}{2}(1-\beta^2)} \]
where $\beta$ is the lepton velocity in the virtual photon rest frame; in the limit of ultrarelativistic lepton daughters this recovers the familiar form $P \propto 1 + \cos^2 \theta$.  These results only depend on the parameters of the virtual photon system and so are universal for AIC events with transversely polarized photons.

\begin{figure}
 \begin{subfigure}[t]{.5\linewidth}
  \includegraphics[width=\linewidth]{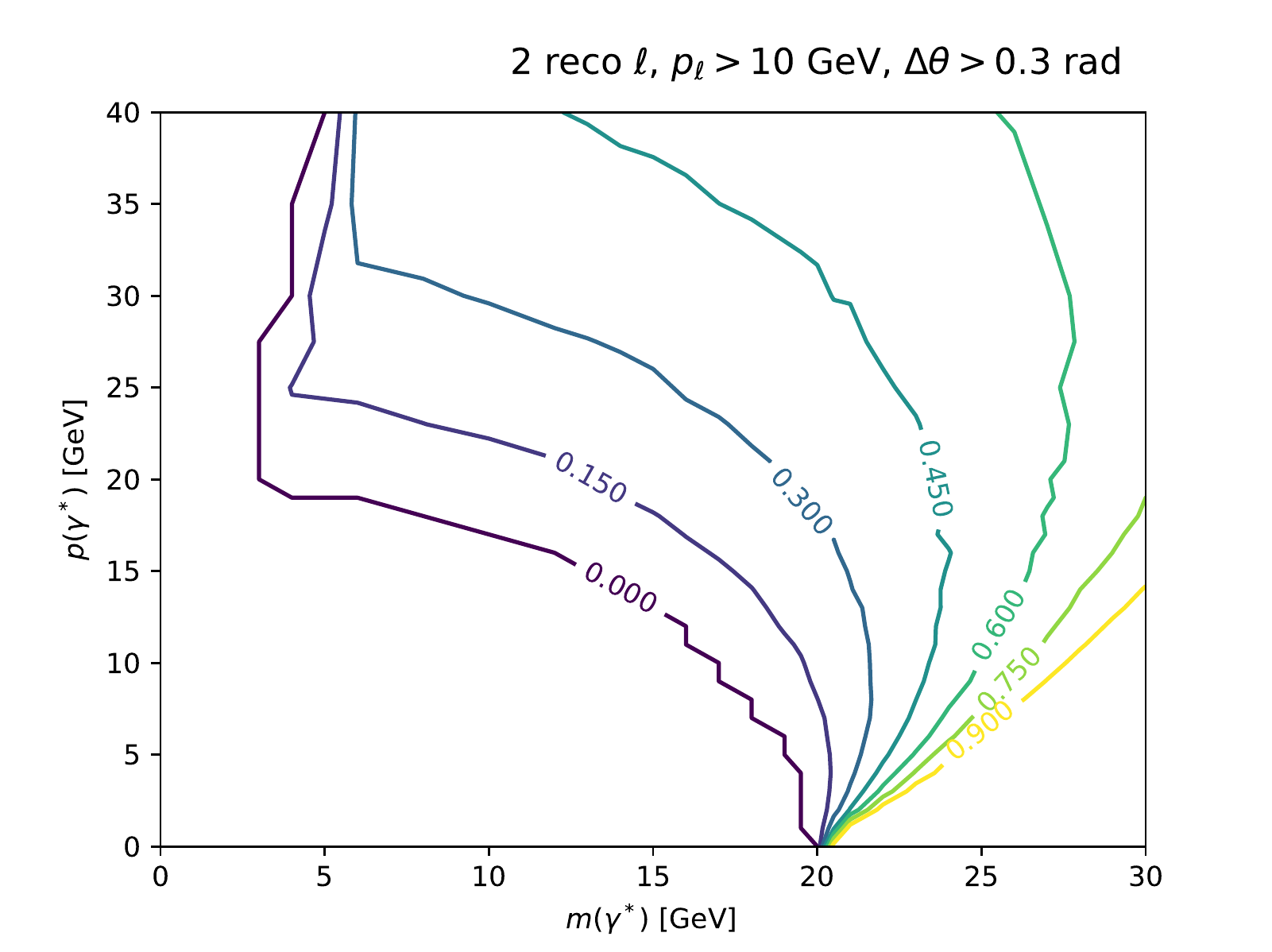}
  \caption{Two lepton selection}
 \end{subfigure}
 \begin{subfigure}[t]{.5\linewidth}
  \includegraphics[width=\linewidth]{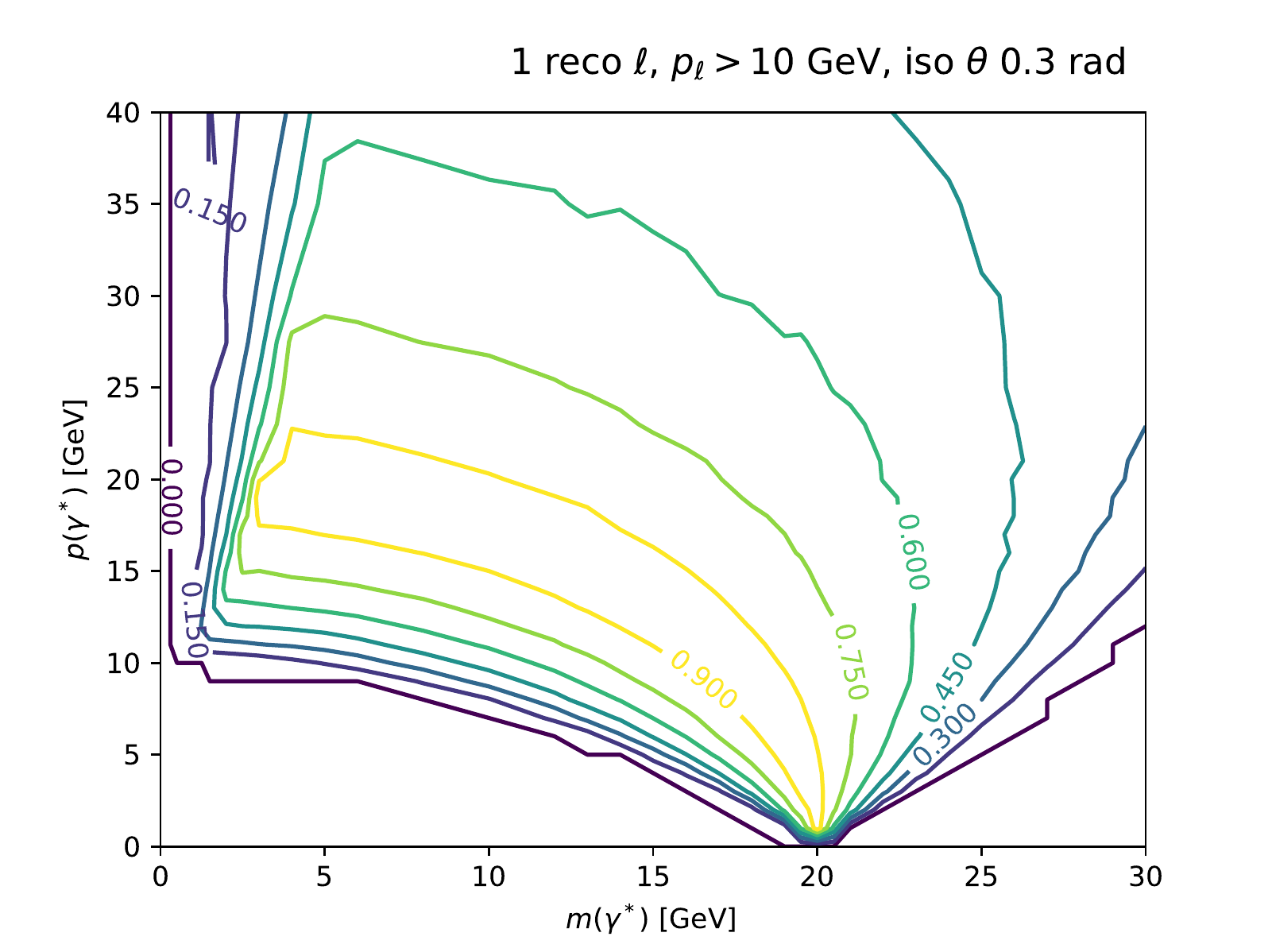}
  \caption{One lepton selection}
 \end{subfigure}
\caption{\label{fig:std_acceptance}Acceptance for two- and one-lepton selections in asymmetric internal conversion $\gamma^* \to \mu\mu$ events.}
\end{figure}

The probability for an event to be chosen under the two selections, as a function of virtual photon mass and momentum, is shown in Figure~\ref{fig:std_acceptance} for the dimuon case.  Several features are evident.  At high $m(\gamma^*)$ and $p(\gamma^*)$, the two plots are complementary, as at least one lepton will have $p > 10$ GeV, and the two selections are disjoint.  At low mass and momentum, no leptons are found at all; they are both too soft. At low mass and high momentum, the lepton separation/isolation requirements suppress the acceptance.  There is a crescent-shaped region of near-zero dilepton acceptance with over 90\% acceptance for the single lepton selection, extending to $m(\gamma^*) \sim 2.5$ GeV; the acceptance is still significant until it is truncated by the isolation cut.  We conclude that an analysis that requires one lepton and vetoes a second will have high sensitivity to these ``lost'' leptons.

\begin{figure}
 \begin{subfigure}[t]{.5\linewidth}
  \includegraphics[width=\linewidth]{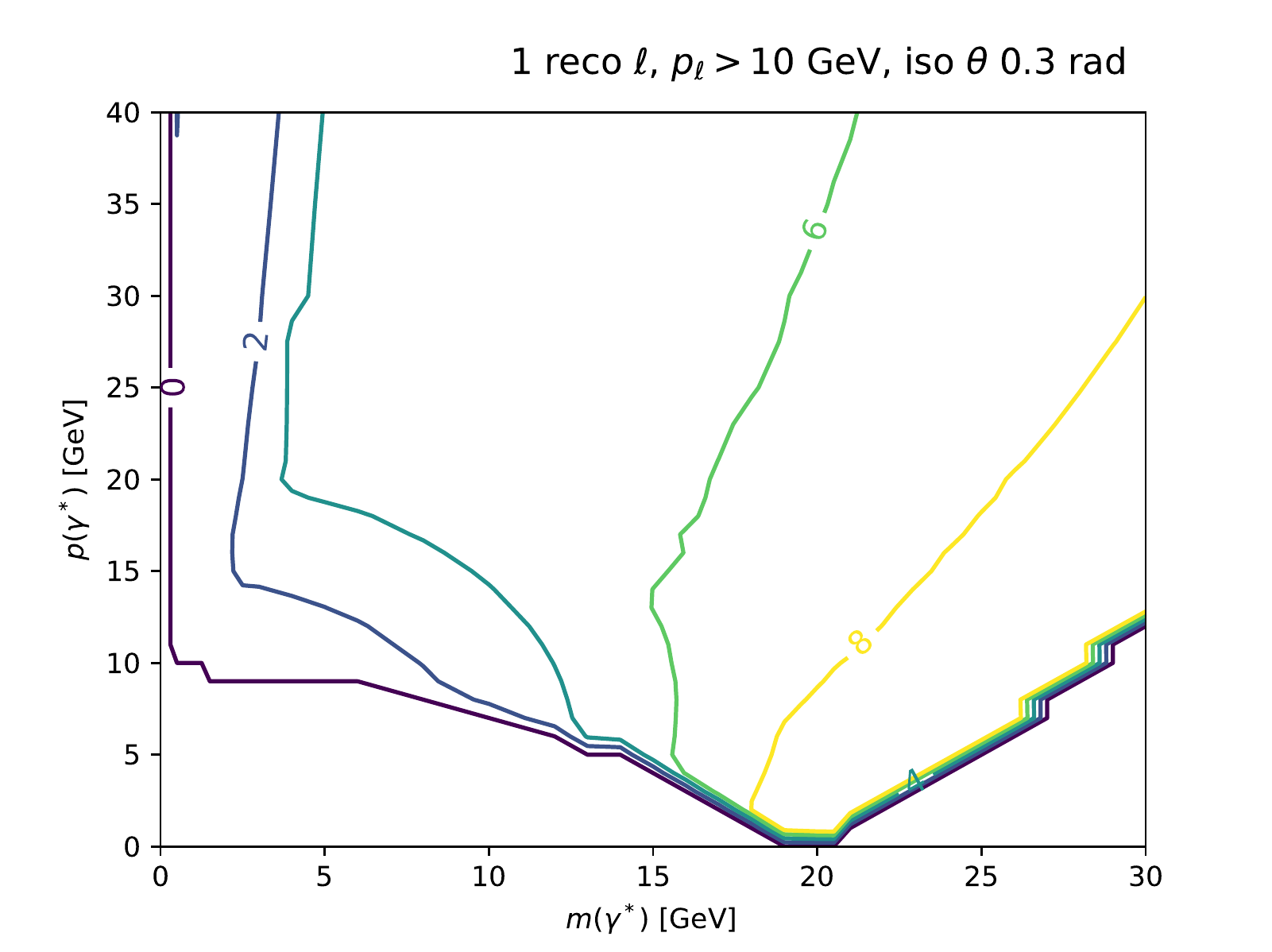}
  \caption{\label{fig:medianp}Median subleading lepton momentum in GeV}
 \end{subfigure}
 \begin{subfigure}[t]{.5\linewidth}
  \includegraphics[width=\linewidth]{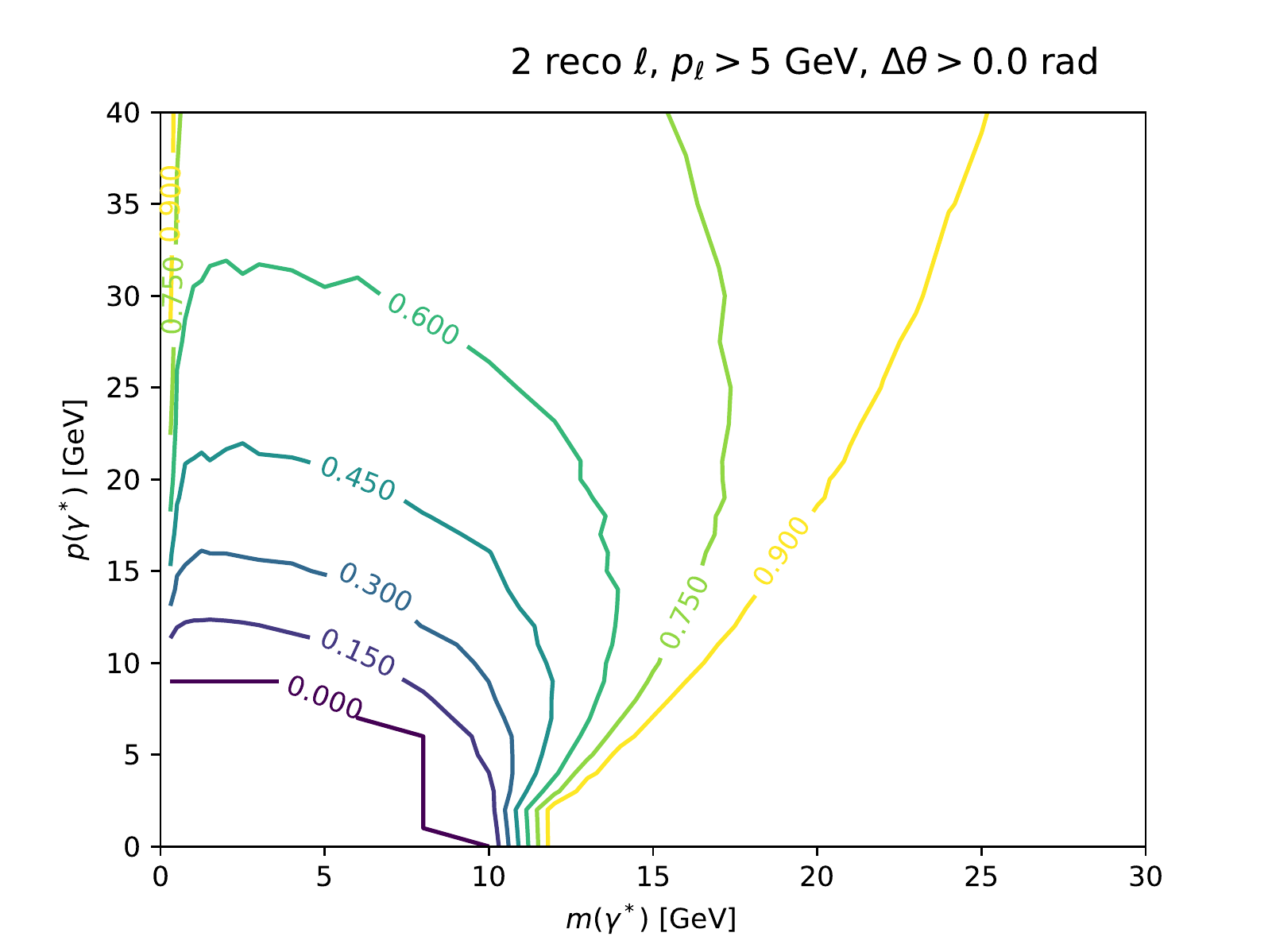}
  \caption{\label{fig:lowp}Two lepton acceptance for $p > 5$ GeV selection}
 \end{subfigure}
\caption{\label{fig:test_acceptance}(Left) median subleading lepton momentum when only one lepton is found with $p > 10$ GeV; (right) acceptance for two-lepton selection when lepton momentum threshold is lowered to 5 GeV and lepton angular separation requirements are removed.}
\end{figure}

We can try to see if there are ways to mitigate this background or to constrain it from data. From Figure~\ref{fig:std_acceptance} it is clear that the acceptance for finding two leptons with $p > 10$ GeV in the region of interest is essentially zero, so we must try other means. First, we ask whether the subthreshold lepton can still be seen, with the leading lepton threshold still at 10 GeV.  Figure~\ref{fig:medianp} shows the median momentum of the subleading lepton when exactly one lepton is found with $p > 10$ GeV; if a subleading lepton can be reconstructed down to 4 GeV momentum, then events down to $m(\gamma^*) = 5$ GeV can be reconstructed in most of the dangerous cresecent region, although the very lowest masses cannot be reached.

We also consider the possibility of lowering the leading lepton threshold to 5 GeV and eliminating the separation cut between two reconstructed leptons in order to improve acceptance at low $m(\gamma^*)$.  The results are shown in Figure~\ref{fig:lowp}.  With this selection, an acceptance of $\gtrsim 30\%$ can be achieved down to dilepton threshold in the dangerous crescent region.  The better matching of the shape of the acceptance curve to the crescent in the latter case, and the ability to reach lower $m(\gamma^*)$, suggest it as a preferred solution to map out the cross section for AIC processes in data.

\section{\label{sec:decay}The Rare Decays $t \to \ell' \nu b \ell\ell$ and $t \to
  q q' b \ell\ell$}
  At 13 TeV, a $\sigma(pp \to t\bar t)$ of 832 pb has been calculated at NNLO+NNLL using the \textsc{Top++} v2.0 program \cite{Czakon:2011xx} by the LHC Top Working Group \cite{topxsec}. 
  By comparison, the $pp \to t\bar tW \to 3\ell 3\nu bb$ ($\ell = e$, $\mu$) cross section at NLO is 6.2 fb \cite{deFlorian:2016spz,Alwall:2014hca}, so a $10^{-5}$ top quark branching fraction to a similar final state could produce comparable yields.
  
Top quark decays are dominated by the CKM-favored two-body decay $t \to Wb$.
Other potential two-body decays are either suppressed by small CKM elements
(for $Wq$) or GIM-suppressed loops ($Zq$ or $Hq$).  The CKM-allowed decay $t
\to WZb$ is very near threshold and the decay width is dominated by
finite-width effects. Ref.~\cite{Altarelli:2000nt} derives a branching fraction $\mathcal{B}(t \to WZb) \sim 2 \times 10^{-6}$ which means this will not compete with $pp \to t\bar t Z$ or $tZ$ production. The decays $t \to Wb g$ and $t \to Wb\gamma$ are significant but generally treated as FSR corrections to $t \to Wb$ using QCD and QED Monte Carlo showering algorithms.

The four-body decay $t \to Wb \ell^+\ell^-$ was considered in Ref.~\cite{Quintero:2014lqa} and found to give a branching ratio $R = \Gamma(t \to Wb e^+e^-)/\Gamma(t \to Wb) = 6.3 \times 10^{-6}$ for $m(\gamma^*)^2 > 20$ GeV$^2$, and the same for the dimuon decay (all at leading order).  This is clearly in the regime discussed above where this decay might compete with processes such as $t\bar tW$, considering in particular that one expects significant cross section below this $m(\gamma^*)$ cutoff of $\approx 4.5$ GeV.  A leading order calculation suffers from the pole of the photon propagator, especially for the dielectron mode where the branching ratio to the $t \to Wb$ mode exceeds unity with an $m(\gamma^*)^2$ cut less than $10^{-3}$ GeV$^2$, which is still above dielectron threshold.  One expects these to be compensated by virtual corrections in a higher order calculation and, indeed, the best strategy for for dealing with these decays is most likely a resummed calculation or parton shower.  Nevertheless, it is instructive to study the fixed leading order calculation.

We use \textsc{MadGraph5\_aMC@NLO} \cite{Alwall:2014hca} (hereafter \textsc{MG5\_aMC}) version 2.5.4, running in leading order mode, to study the inclusive processes $t \to \ell' \nu b \ell\ell$ and $t \to q q' b \ell \ell$.  We use $m_t = 173$ GeV and use non-zero lepton masses in the calculation. First, we compute the ratio $R =\Gamma(t \to Wb e^+ e^-)/\Gamma(t \to Wb)$ for $m(\gamma^*)^2> 20$ GeV$^2$, to compare with Ref.~\cite{Quintero:2014lqa}.  Our result is $5.8 \times 10^{-6}$, in reasonable agreement with the earlier calculation. We find that increasing the top mass by 0.5 GeV results in an increase in the ratio of $0.08 \times 10^{-6}$, indicating good stability with the top quark mass choice.

In the calculation of Ref.~\cite{Quintero:2014lqa} the possibility of the virtual photon being radiated from the daughters of the $W$ boson is not considered.  Considering the full set of possible diagrams for $t \to (\ell'\nu, q q')b e^+ e^-$, we compute a somewhat larger $R$ of $1.1 \times 10^{-5}$ for the same cutoff $m(\ell^+\ell^-)^2 > 20$ GeV$^2$.  (In the case of the three leptons having the same flavor, the $m(\ell\ell)$ threshold is applied on both opposite sign pairs.) We find that $\Gamma(qq'b e^+ e^-)/\Gamma(\ell'\nu e^+ e^-) = 1.2$ summing over all lepton flavors $\ell'$, which is much smaller than the equivalent ratio $\Gamma(qq' b)/\Gamma(\ell' \nu b) \approx 2$.  This feature seems to be robust to various generation options and may arise from a) the smaller charges of the quarks compared to the charged leptons and b) interference between diagrams involving $W$-daughter radiation and other diagrams. In Section~\ref{sec:simulation} we focus on the $\ell'\nu b \ell^+ \ell^-$ case and so are not sensitive to potential complications in $W$ hadronic decays.

For the simulations of Section~\ref{sec:simulation} we choose a lower $m(\ell^+\ell^-)$ threshold, 1 GeV. This allows us to fully populate the high-acceptance region for AIC events. The corresponding $R$ is $4.9 \times 10^{-5}$. The simulation of $t\bar t$ events with a rare decay is performed completely at LO in \textsc{MG5\_aMC}, with the rare decays simulated as a 5-body decay.  
Although this simulation will miss features such as  recoil of the $t\bar t$ system against hard additional jets, it will illustrate the core physics.  We strongly emphasize that higher-order corrections are expected to induce large corrections and that in the leading order calculation there is a large additional cross section at $m(\gamma^*) < 1$ GeV. This $R$ value should therefore be treated as purely indicative.

\section{\label{sec:simulation}Simulation of Rare Top Decay Impact on $3\ell + b$ Searches}
To give a practical picture of the potential impact of these rare top decays
on an actual search,
we consider a simplified $3\ell + b$ search with a $Z$-veto.  This final state is sensitive to
$t\bar t W$ and $t\bar t H$.

The sensitivity of such an analysis to asymmetric conversions depends strongly
on detector response and analysis choices.  We pass events generated in
\textsc{MG5\_aMC} at 13 TeV and showered by \textsc{Pythia 8} \cite{Pythia8,Sjostrand:2006za} to the generic detector fast simulation program
\textsc{Delphes} \cite{deFavereau:2013fsa}, which was configured to approximate the characteristics of the CMS
detector. We generate three different processes.  The $t\bar t$ rare decays produce the final state $\ell_1^+ \nu_1 b \ell_2^- \bar \nu_2 \bar b \ell_3^+ \ell_3^-$, where all leptons are either $e$ or $\mu$ and the $\ell_3^+ \ell_3^-$ pair can come from the decay of either $t$ or $\bar t$.  The $pp \to t\bar t W$ process is generated with subsequent decay of all $W$ bosons to a lepton and a neutrino, where here leptons include $e$, $\mu$, and $\tau$.  The $pp \to t\bar t H$ process is generated with subsequent $H \to WW^*$ decay; three of four $W$ bosons are required to decay to a lepton and neutrino (including $e$, $\mu$, and $\tau$) and the fourth is required to decay to $qq'$.  For $t\bar t$ we assign a cross section of 10.9 fb using a branching fraction $\mathcal{B}(t \to \ell'\nu b \ell\ell)$ computed using \textsc{MG5\_aMC} and a total inclusive $t\bar t$ cross section of 832 fb. For $t \bar t W$ and $t \bar t H$ we assign cross sections of 20.8 and 10.3 fb, respectively, derived from production cross sections and decay branching fractions in the LHC Higgs Cross Section Working Group Yellow Report 4 \cite{deFlorian:2016spz} and the Particle Data Group averages \cite{Olive:2016xmw}.  All processes are generated at leading order with no additional partons produced in the matrix element.

The default \textsc{Delphes} CMS simulation card was altered to use anti-$k_T$
$R=0.4$ jets with a $p_T$ threshold of 25 GeV, and to require that all reconstructed leptons satisfy $p_T > 10$
GeV.  Leptons are required to be isolated; a particle flow algorithm is used
to classify additional activity in a cone of $\Delta R \equiv \sqrt{\Delta
  \eta^2 + \Delta \phi^2} = 0.3$ around the lepton, and the total transverse
energy of particles with $p_T > 0.5$ GeV in this cone is required to be less
than 10\% of the lepton $p_T$.  To remove quarkonia decaying to muon pairs, typical analyses will require that the invariant mass of any opposite sign same flavor lepton pairs to exceed 12 GeV, which will also have the advantage of removing low mass internal conversion events where both leptons are reconstructed.

We consider three possible analysis selection scenarios:
\begin{enumerate}
 \item exactly three reconstructed leptons with $p_{T} > 10$ GeV, no further selection on jets;
 \item exactly three reconstructed leptons with $p_{T} > 20$ GeV, no further selection on jets;
 \item exactly three reconstructed leptons with $p_{T} > 20$ GeV, with $\ge 4$ jets required.
\end{enumerate}
These are motivated by different potential analysis targets.  Scenario 1 illustrates generic features of the different processes and might be used as a preselection for a multivariate discriminant analysis that seeks as much acceptance as possible. Scenario 2 attempts to reduce the impact of the $t\bar t$ rare decay background by raising lepton momentum thresholds (this would also be motivated by reducing the contamination from leptons from $b$- and $c$-hadron decays), and would be a starting point for a $t\bar t W$ selection.  Scenario 3 is closest to a selection targeting $t\bar t H$, requiring high jet multiplicity to suppress the $t \bar t W$ contribution.  In all cases triggering should be very efficient and is not considered.

We make no specific selection on the number of $b$-tagged jets. All three samples ($t\bar t$ rare decay, $t\bar t W$, $t\bar t H$) have virtually identical distributions of the number of reconstructed $b$-tagged jets, consistent with the tagging efficiency implemented in the \textsc{Delphes} card.  To improve statistical yields for comparisons we do not cut on this variable.

\begin{figure}
 \begin{subfigure}[t]{.5\linewidth}
  \includegraphics[width=\linewidth]{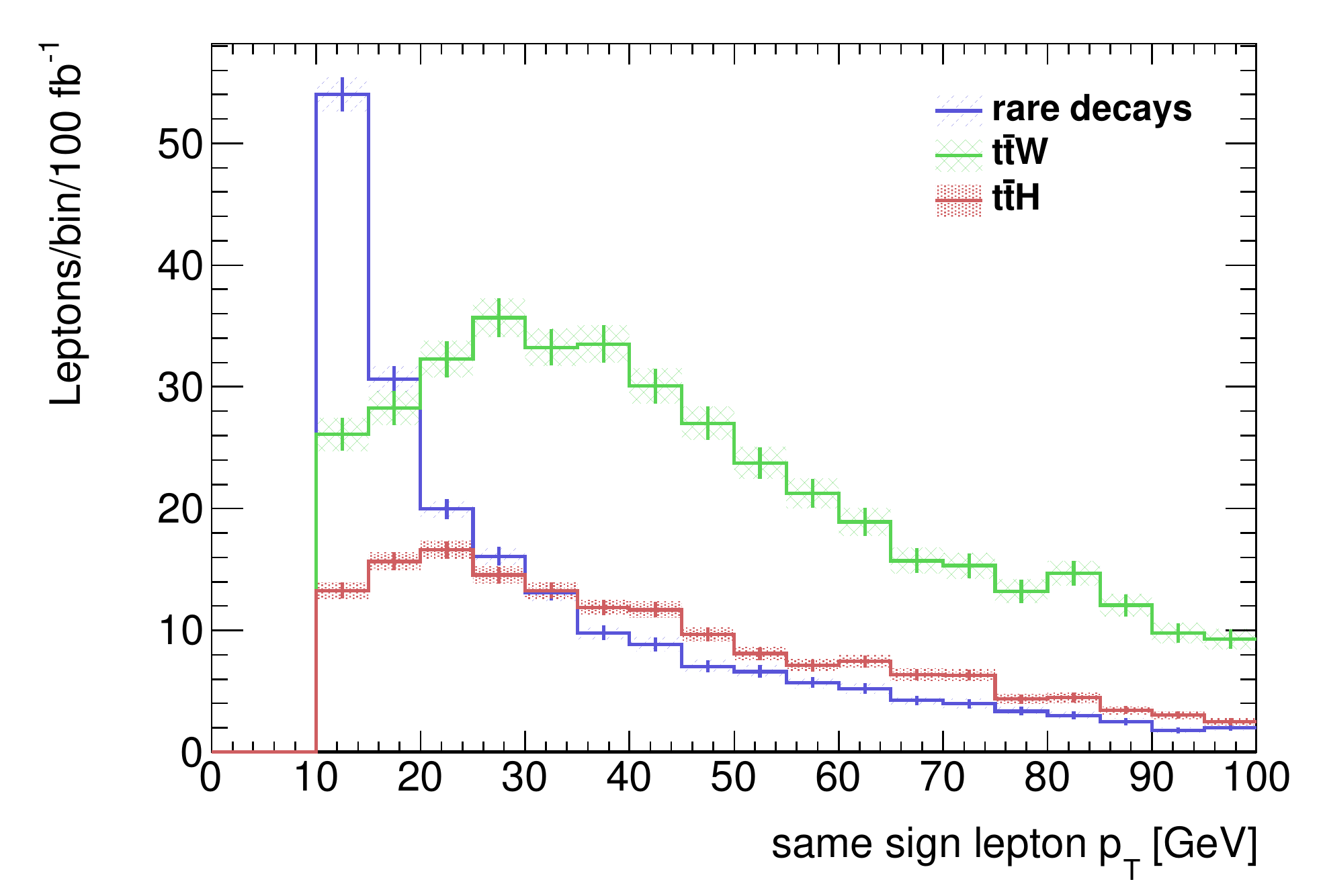}
  \caption{\label{fig:00sspt}}
 \end{subfigure}
 \begin{subfigure}[t]{.5\linewidth}
  \includegraphics[width=\linewidth]{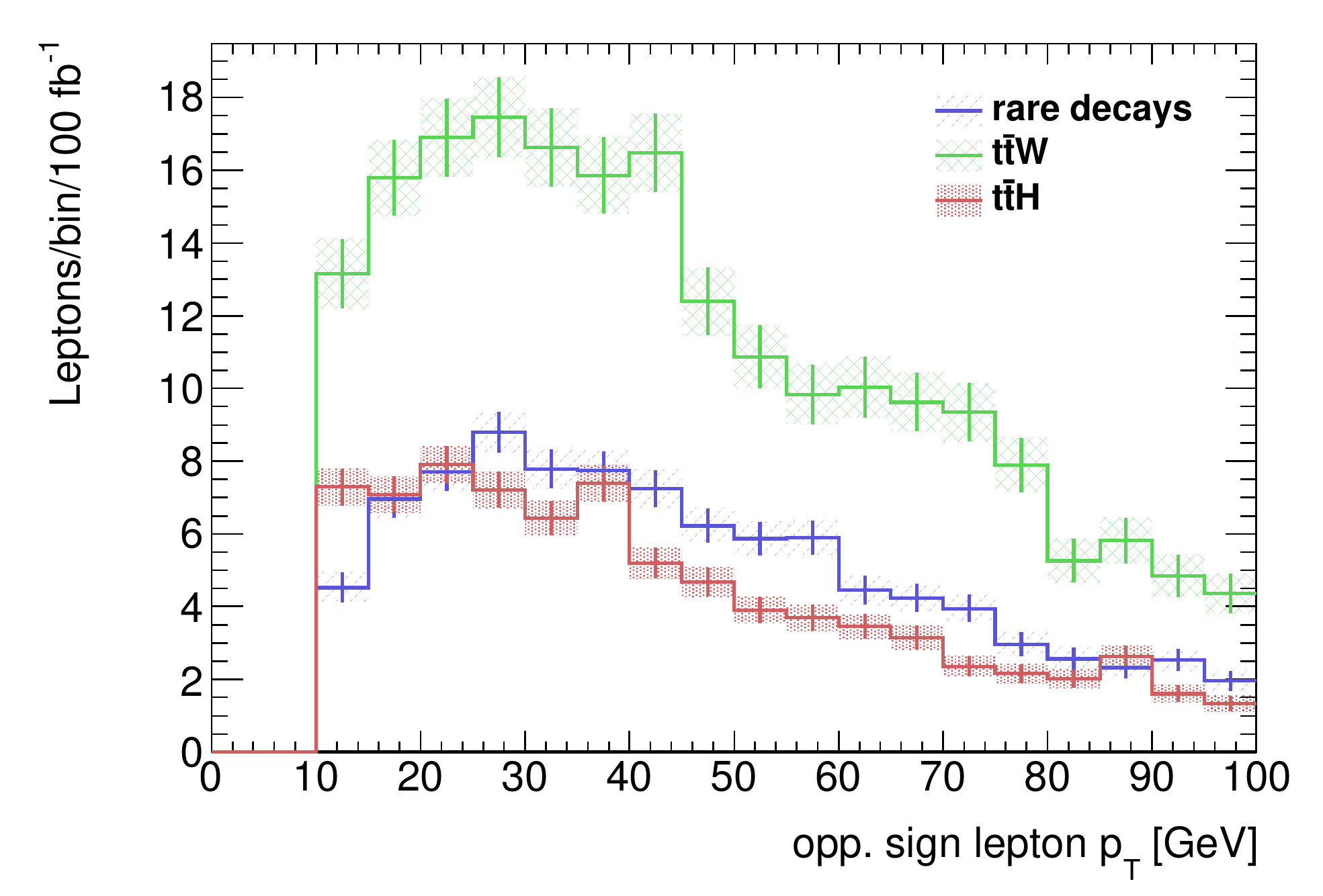}
  \caption{\label{fig:00ospt}}
 \end{subfigure}
 \begin{subfigure}[t]{.5\linewidth}
  \includegraphics[width=\linewidth]{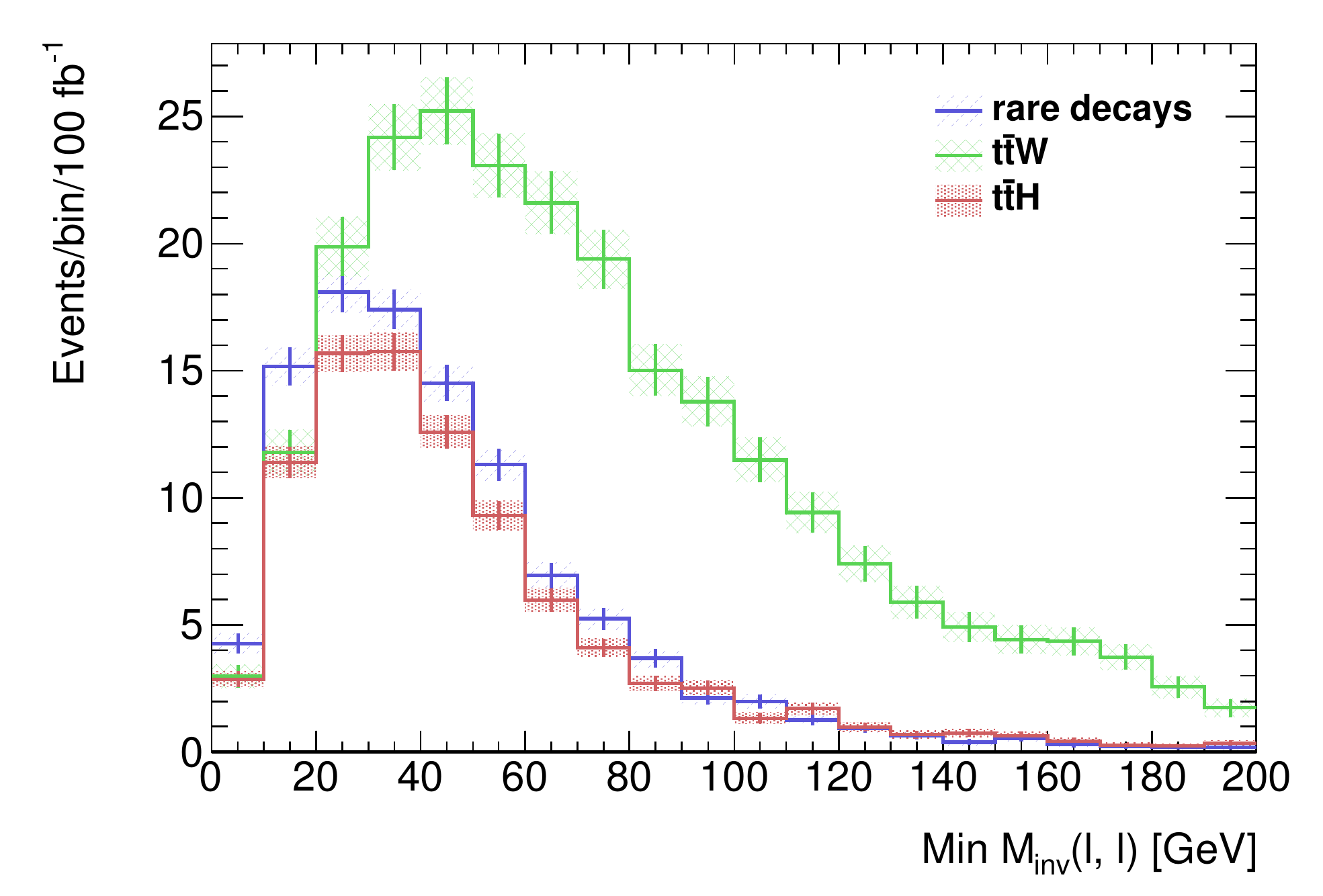}
  \caption{\label{fig:00mllmin}}
 \end{subfigure}
 \begin{subfigure}[t]{.5\linewidth}
  \includegraphics[width=\linewidth]{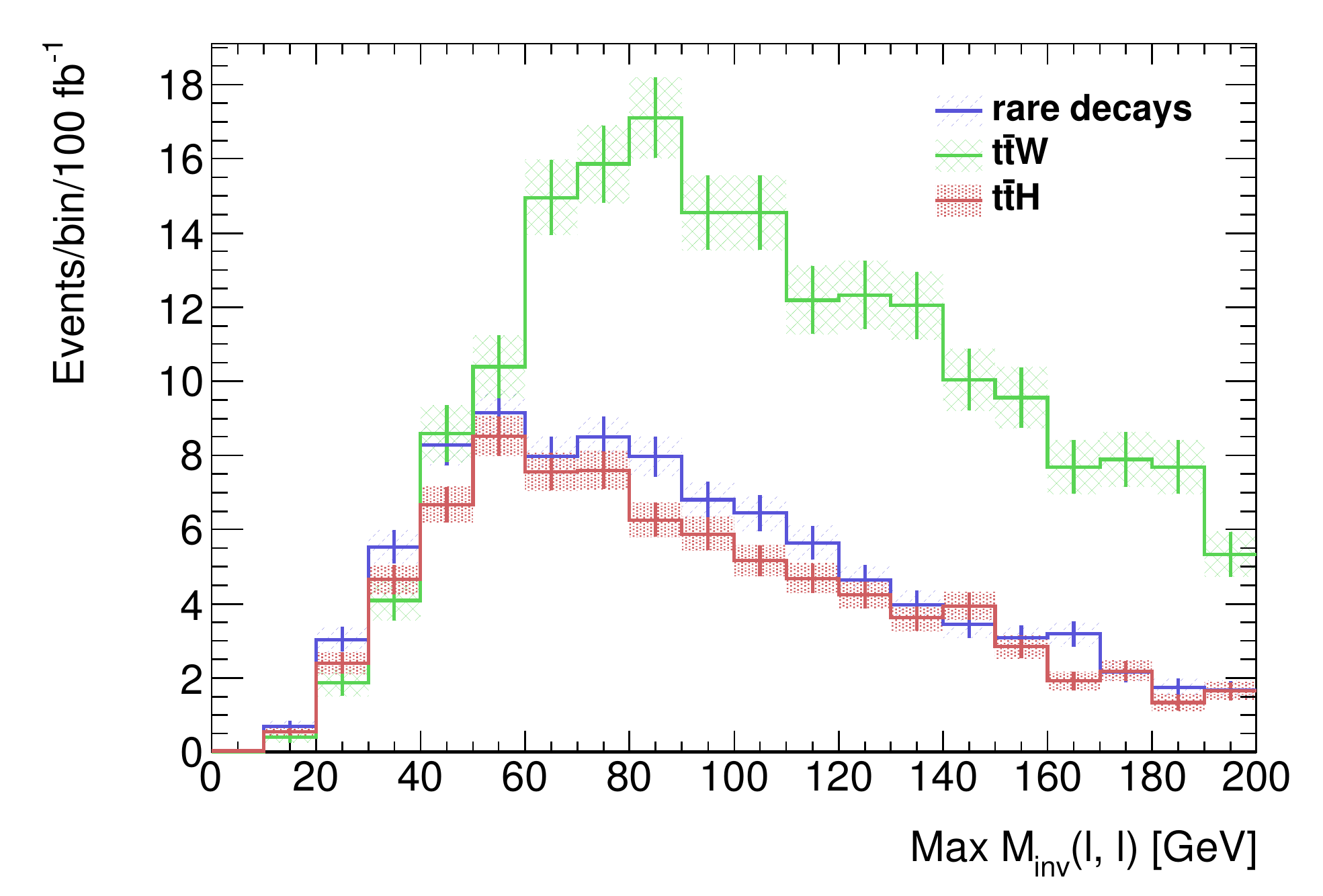}
  \caption{\label{fig:00mllmax}}
 \end{subfigure}
 \begin{subfigure}[t]{.5\linewidth}
  \includegraphics[width=\linewidth]{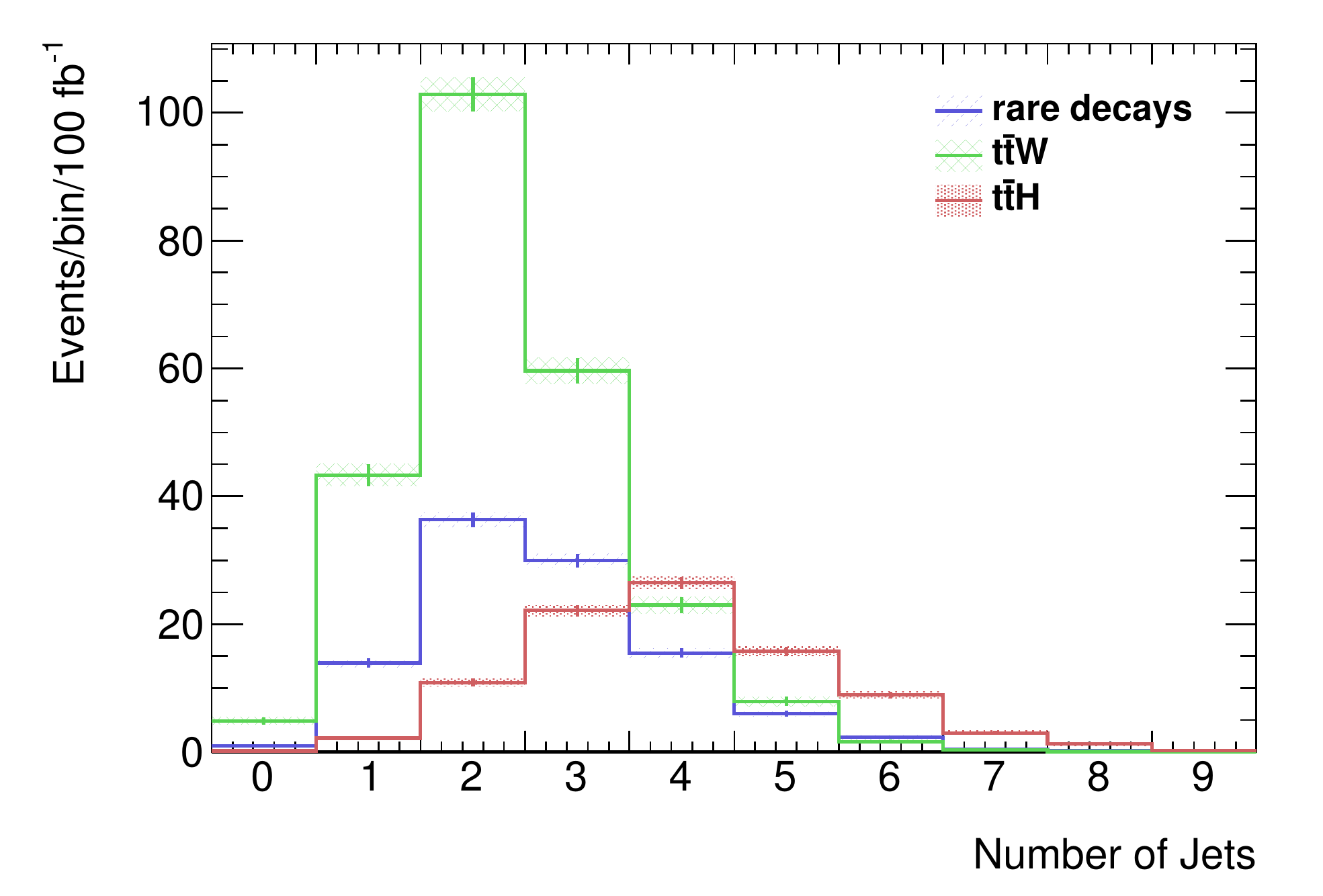}
  \caption{\label{fig:00njet}}
 \end{subfigure}
 \begin{subfigure}[t]{.5\linewidth}
  \includegraphics[width=\linewidth]{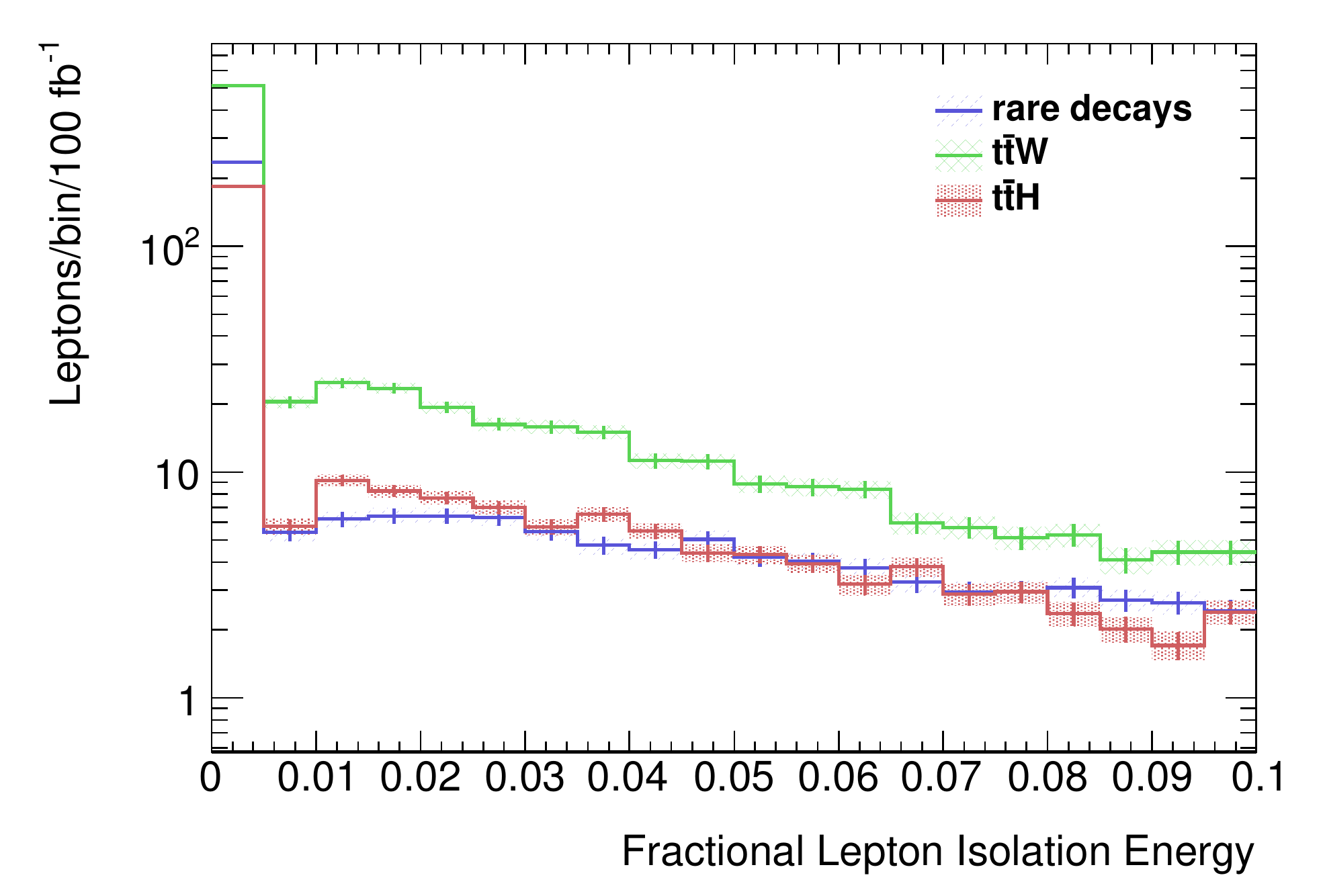}
  \caption{\label{fig:00iso}}
 \end{subfigure}
  \caption{\label{fig:scen1}Plots for various variables in the Scenario 1 selection: a) $p_T$ of the two selected leptons of the same charge; b) $p_T$ of the selected lepton of opposite charge to the other two; c) the smaller of the two possible opposite sign dilepton masses $M(\ell^+\ell^-)$; d) the larger of the two possible opposite sign dilepton masses $M(\ell^+\ell^-)$; e) number of reconstructed jets with $p_T > 25$ GeV; f) ratio of isolation $p_T$ in a cone of 0.3 around the lepton to the lepton $p_T$. Yields are scaled to 100 fb$^{-1}$. Error bands correspond to statistical uncertainties in the simulated samples.}
\end{figure}

\begin{figure}
 \begin{subfigure}[t]{.5\linewidth}
  \includegraphics[width=\linewidth]{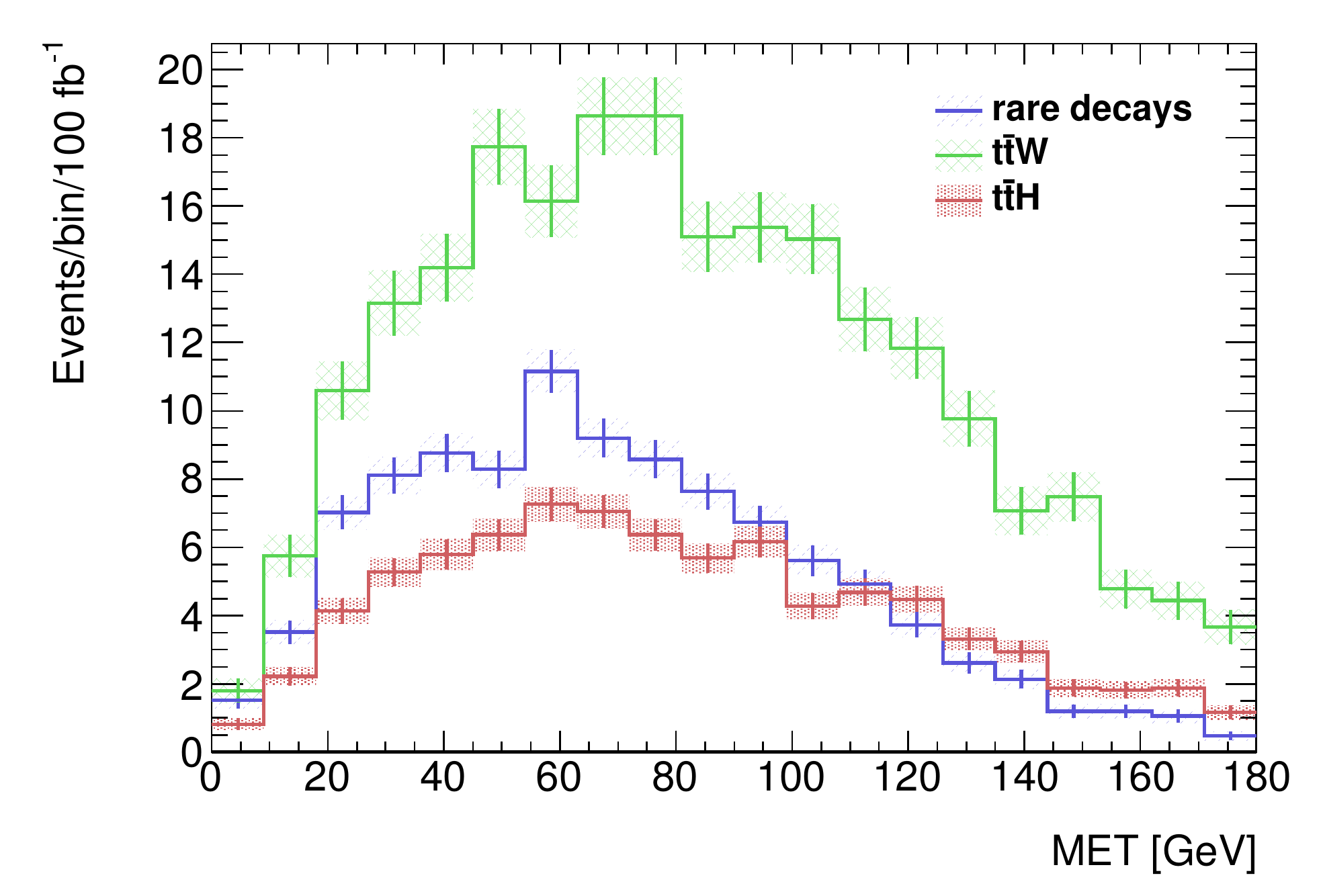}
  \caption{\label{fig:00met}}
 \end{subfigure}
 \begin{subfigure}[t]{.5\linewidth}
  \includegraphics[width=\linewidth]{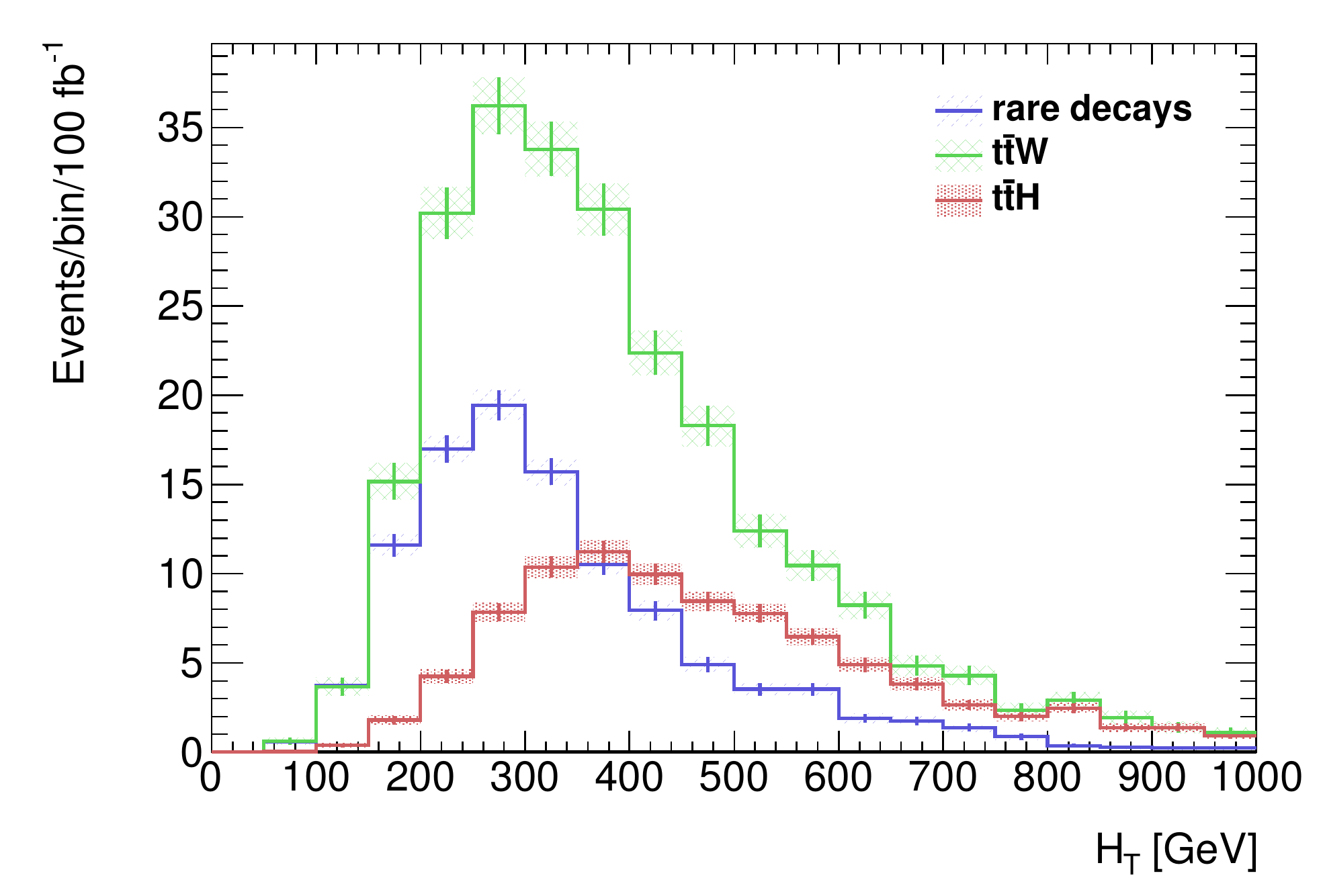}
  \caption{\label{fig:00ht}}
 \end{subfigure}
 \caption{\label{fig:scen1b}Additional plots for various variables in the Scenario 1 selection: a) missing transverse energy; b) scalar sum of transverse momenta $H_T$. Error bands correspond to statistical uncertainties in the simulated samples.}
\end{figure}

The results for Scenario 1 are shown in Figures~\ref{fig:scen1} and \ref{fig:scen1b}.  Figure~\ref{fig:00sspt} shows the transverse momentum spectrum for the two leptons with the same charge.  In the case of the $t\bar t$ rare decay with one AIC lepton lost, the AIC lepton that is found will necessarily be one of these same-charge leptons.  The large contribution of lower-momentum leptons in the rare decay is clear.  Figure~\ref{fig:00ospt} shows the same for the one lepton of opposite charge to the other two; this is expected to never come from a conversion, and indeed no low-$p_T$ peak is seen.  (In fact the $t\bar t H$ process has a larger fractional contribution from the lowest $p_T$ leptons, due to the decay of the low mass offshell $W^*$.)  Figures~\ref{fig:00mllmin} and \ref{fig:00mllmax} show the distributions for the smaller and larger of the two possible invariant masses formed between opposite charge leptons in each event.  In both cases the $t\bar t$ rare decay distribution looks similar to that of $t \bar t H$ and softer than the spectrum for $t \bar t W$. The number of jets is shown in Figure~\ref{fig:00njet}; this peaks at 2 for the $t\bar t$ rare decays.  The spectrum for $t\bar t W$ is similar, and $t \bar t H$ peaks towards higher multiplicity.  Figure~\ref{fig:00iso} shows the ratio of the isolation $p_T$ to the lepton $p_T$.  Although this shows some slope difference between the rare decay and $t\bar t W$/$t \bar t H$, this appears to arise primarily from the different spectrum of the lepton $p_T$ which appears in the denominator, rather than any difference in the isolation energy itself.  Finally, Figures~\ref{fig:00met} and \ref{fig:00ht} show the missing transverse energy and scalar sum of transverse momenta $H_T$; the three processes are not dramatically different in these distributions, although $t\bar t H$ has a larger $H_T$ consistent with having more jets in the final state.

\begin{figure}
 \begin{subfigure}[t]{.5\linewidth}
  \includegraphics[width=\linewidth]{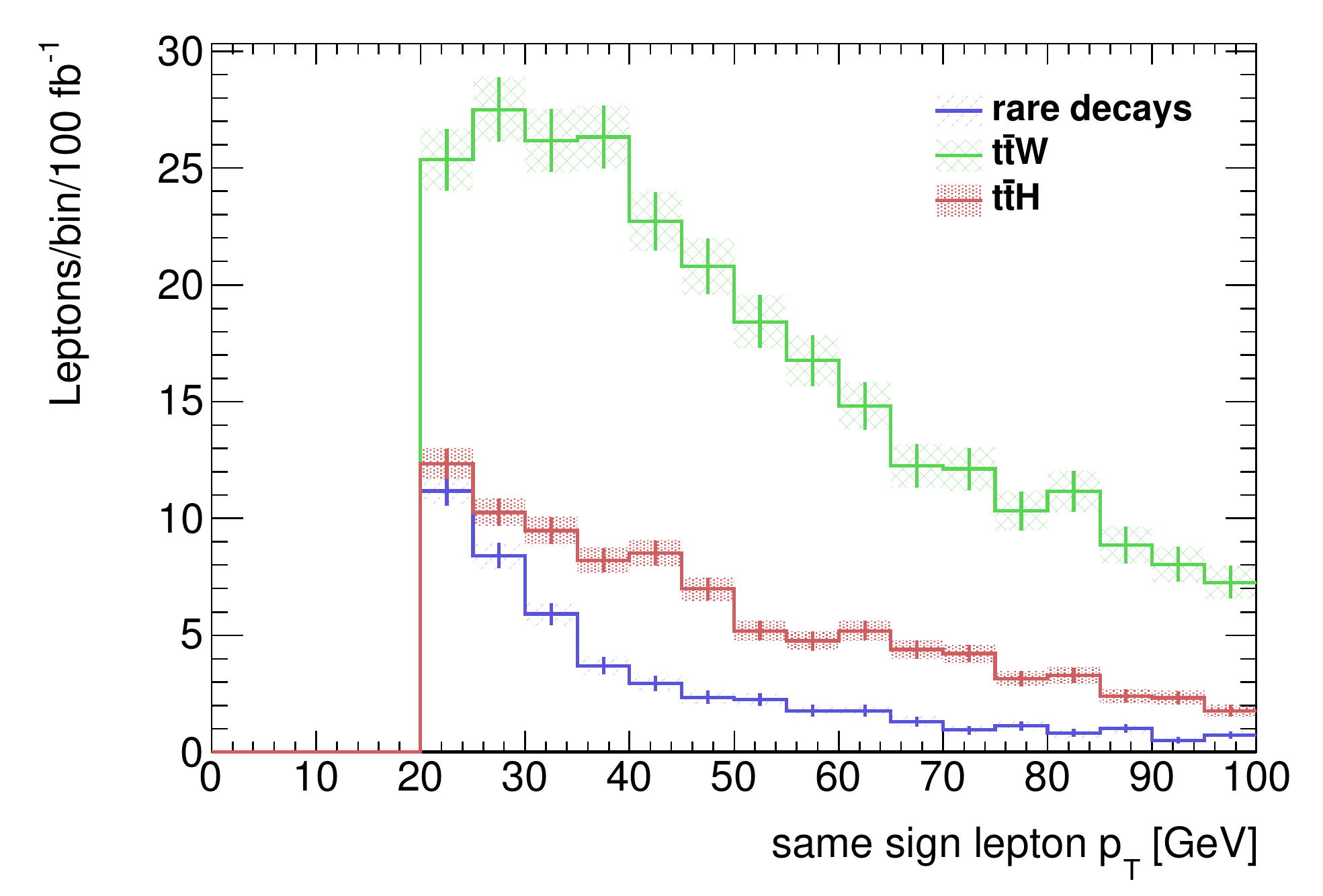}
  \caption{\label{fig:01sspt}}
 \end{subfigure}
 \begin{subfigure}[t]{.5\linewidth}
  \includegraphics[width=\linewidth]{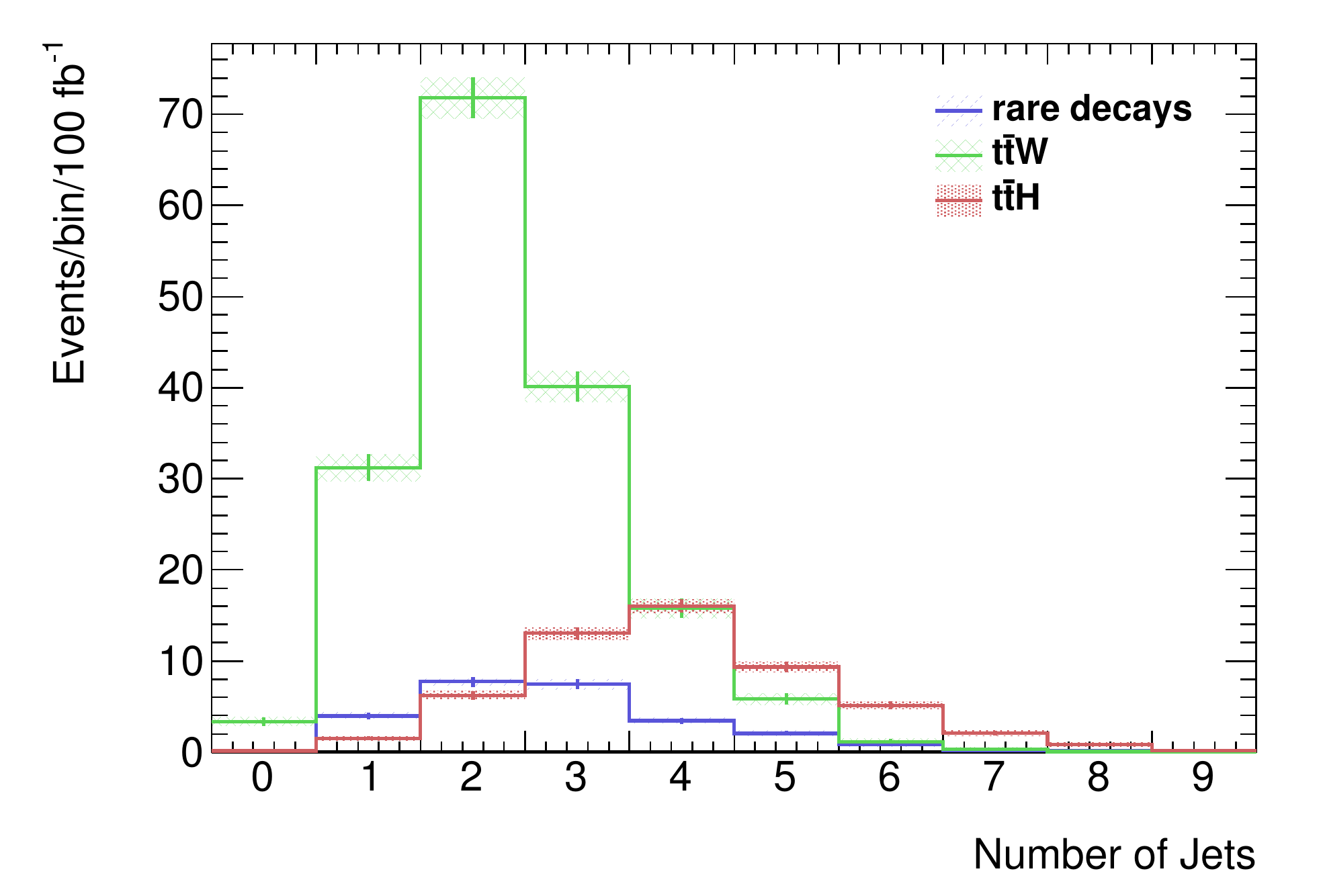}
  \caption{\label{fig:01njet}}
 \end{subfigure}
 \begin{subfigure}[t]{.5\linewidth}
  \includegraphics[width=\linewidth]{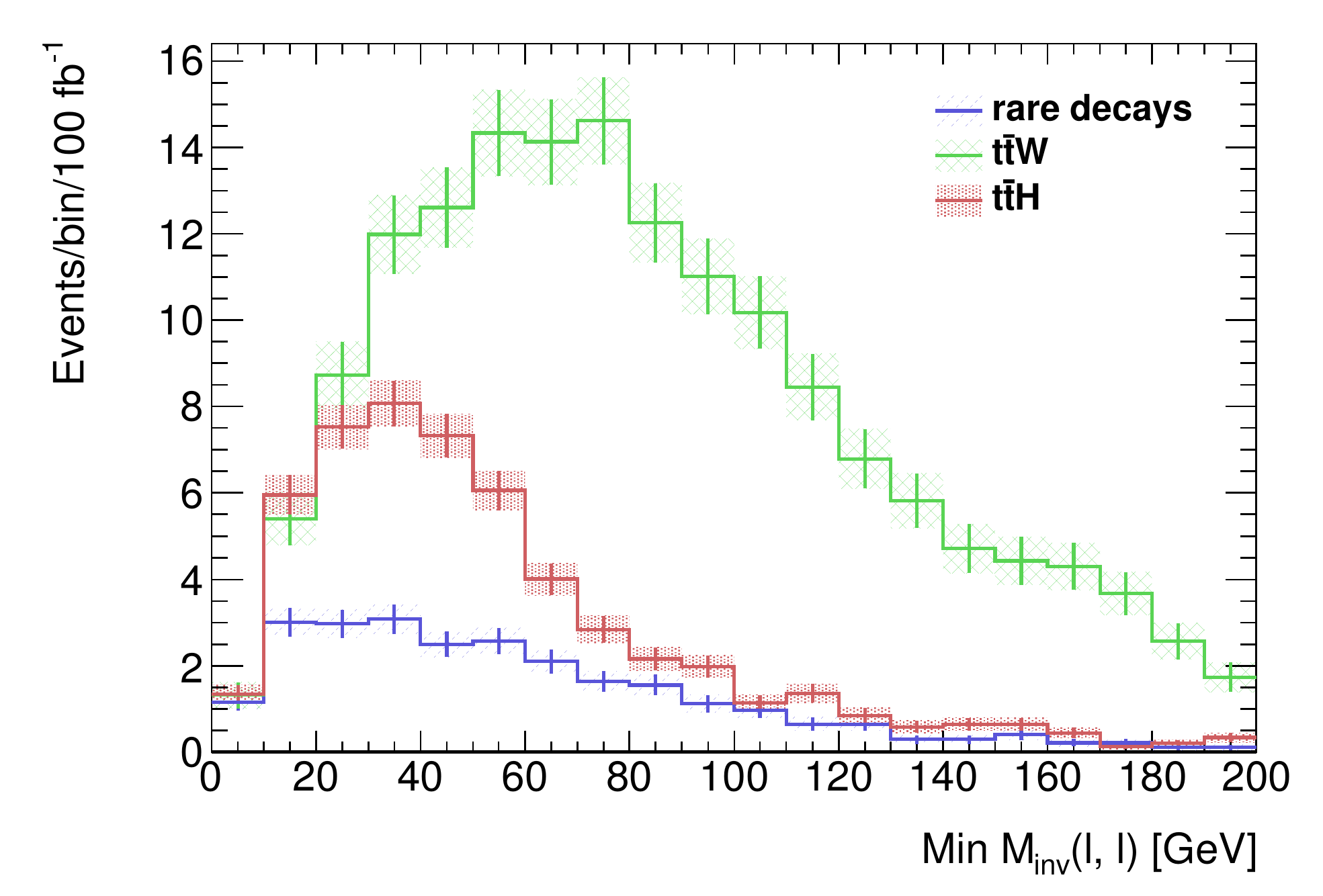}
  \caption{\label{fig:01mllmin}}
 \end{subfigure}
 \begin{subfigure}[t]{.5\linewidth}
  \includegraphics[width=\linewidth]{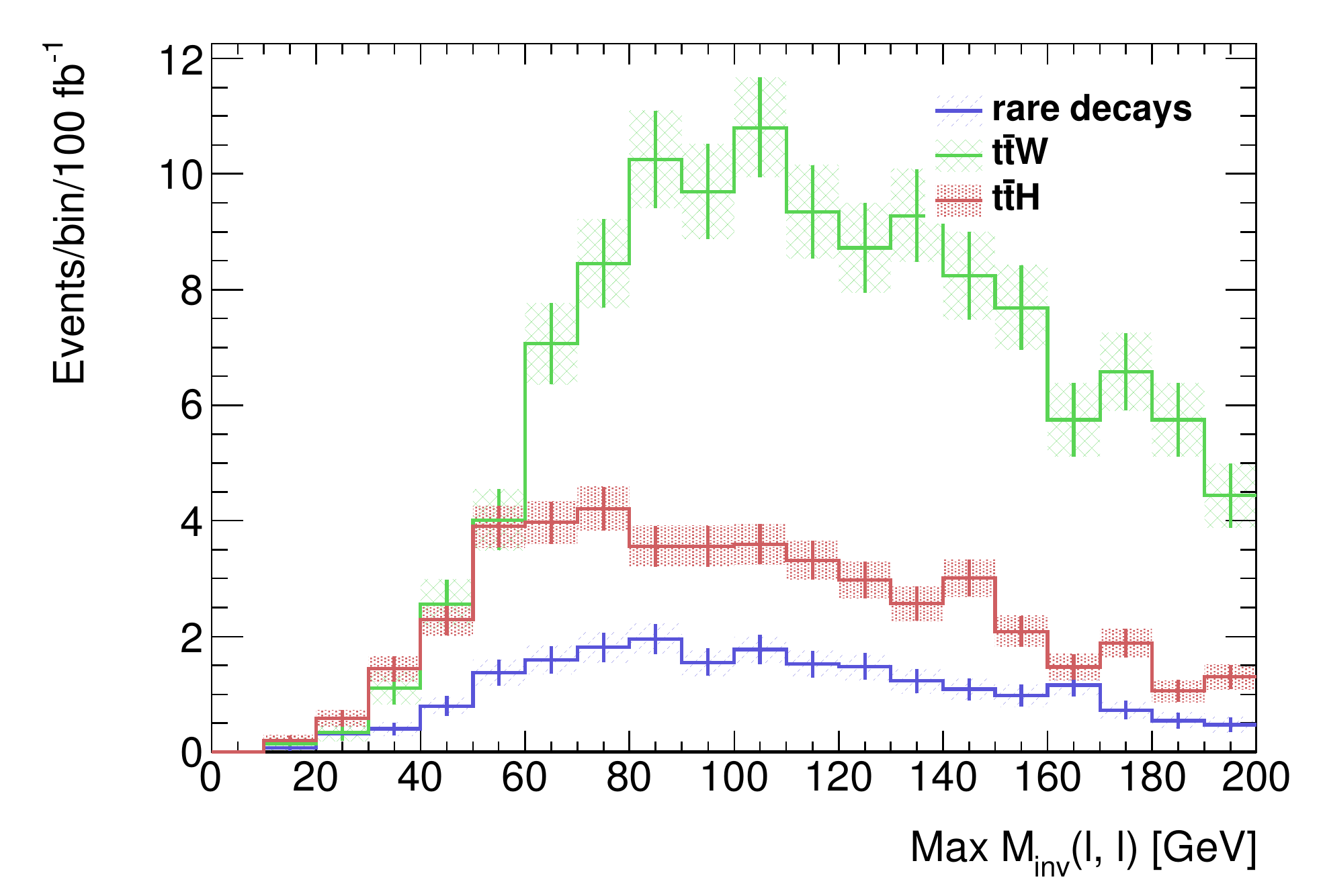}
  \caption{\label{fig:01mllmax}}
 \end{subfigure}
 \begin{subfigure}[t]{.5\linewidth}
  \includegraphics[width=\linewidth]{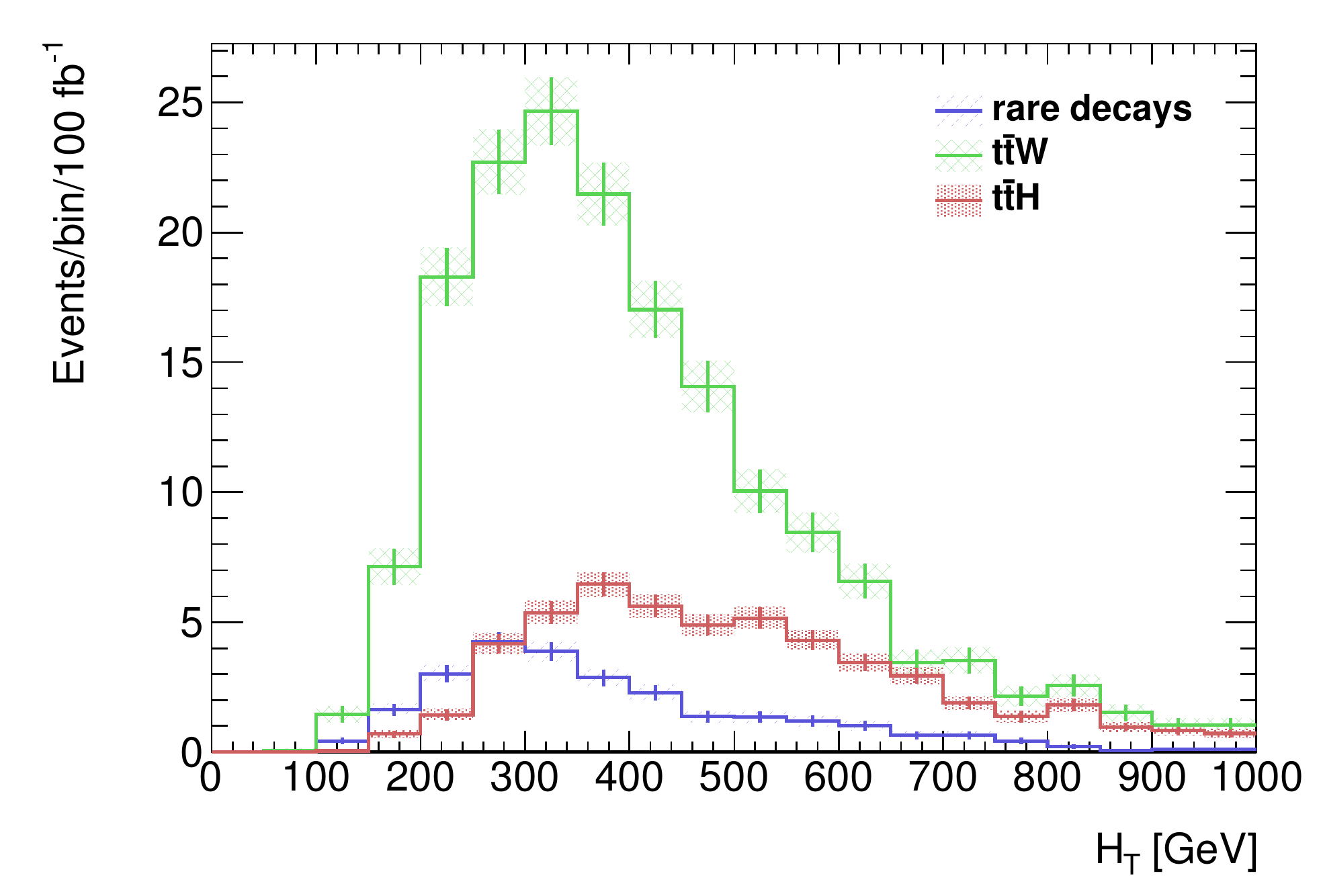}
  \caption{\label{fig:01ht}}
 \end{subfigure}
 \begin{subfigure}[t]{.5\linewidth}
  \includegraphics[width=\linewidth]{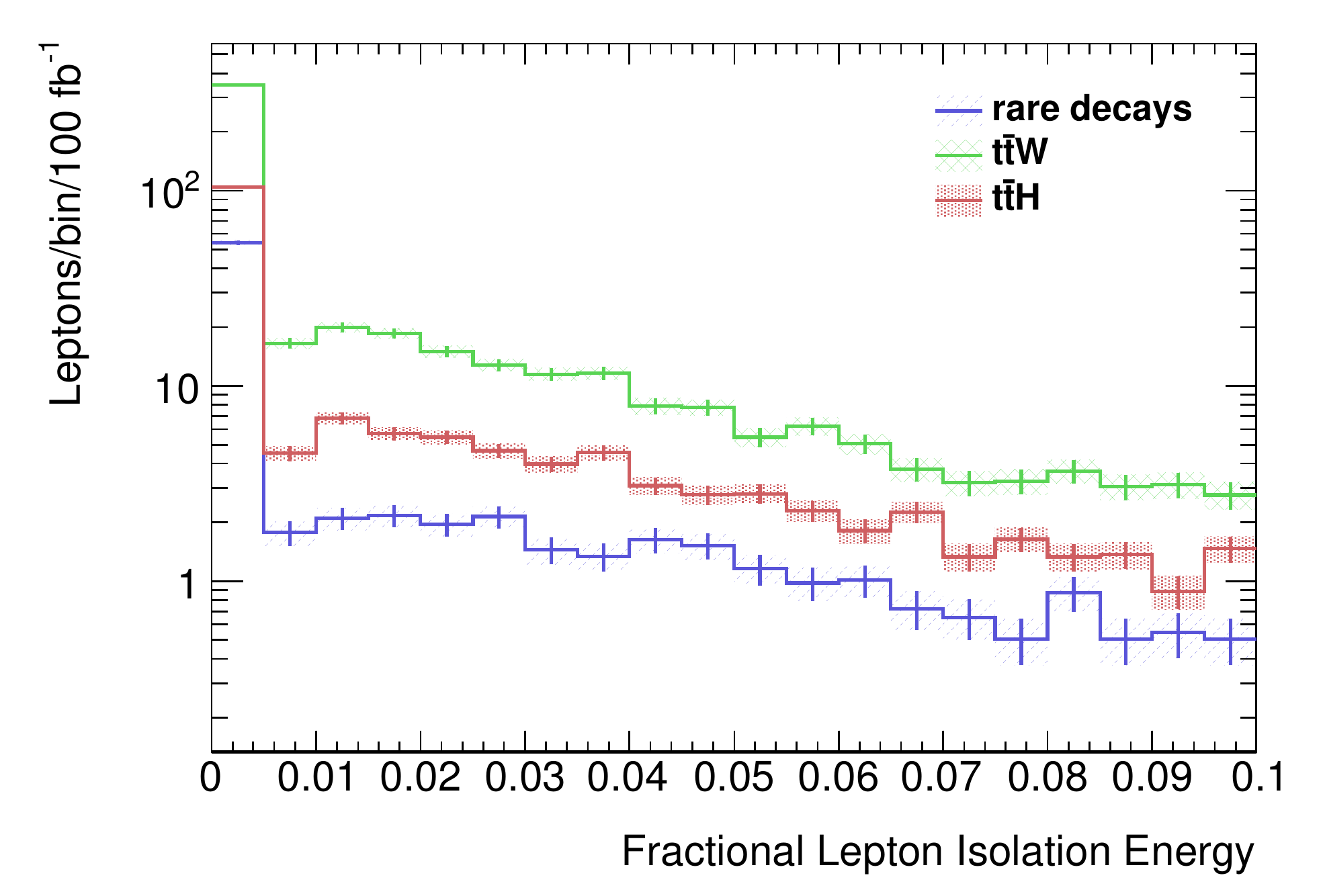}
  \caption{\label{fig:01iso}}
 \end{subfigure}
 \caption{\label{fig:scen2} Plots for various variables in the Scenario 2 selection: a) $p_T$ of the two selected leptons of the same charge; b) number of reconstructed jets with $p_T > 25$ GeV; c) the smaller of the two possible opposite sign dilepton masses $M(\ell^+\ell^-)$; d) the larger of the two possible opposite sign dilepton masses $M(\ell^+\ell^-)$; e) scalar sum of transverse momenta $H_T$; f) ratio of isolation $p_T$ in a cone of 0.3 around the lepton to the lepton $p_T$. Yields are scaled to 100 fb$^{-1}$. Error bands correspond to statistical uncertainties in the simulated samples.}
 \end{figure}
 
 Typically, to suppress non-prompt lepton contributions, the $p_T$ threshold on the leptons will be higher than 10 GeV.  In Scenario 2 we apply a uniform 20 GeV cut on all leptons.  (In principle one can do this for the same sign pair only, but the conclusions are the same.)  The results are shown in Figure~\ref{fig:scen2} for certain selected variables.  The $pp \to t\bar t W$ process is now strongly enhanced over both the rare $t\bar t$ decay and $t\bar t H$ production. Figure~\ref{fig:01sspt} shows the $p_T$ of the same-charge leptons.  The distribution shows a kink at $\approx 40$ GeV for the rare decay, above which it has a very similar momentum spectrum to that of $t\bar t H$.  The invariant mass distributions in Figures~\ref{fig:01mllmin} and \ref{fig:01mllmax} look almost identical between the rare decay process and $t\bar t H$; the two processes are distinguished by the number of associated jets and the $H_T$.  Now that the low-$p_T$ spike in the rare decay has been removed, the isolation energy ratio (Figure~\ref{fig:01iso}) looks very similar for all processes.

 \begin{figure}
 \begin{subfigure}[t]{.5\linewidth}
  \includegraphics[width=\linewidth]{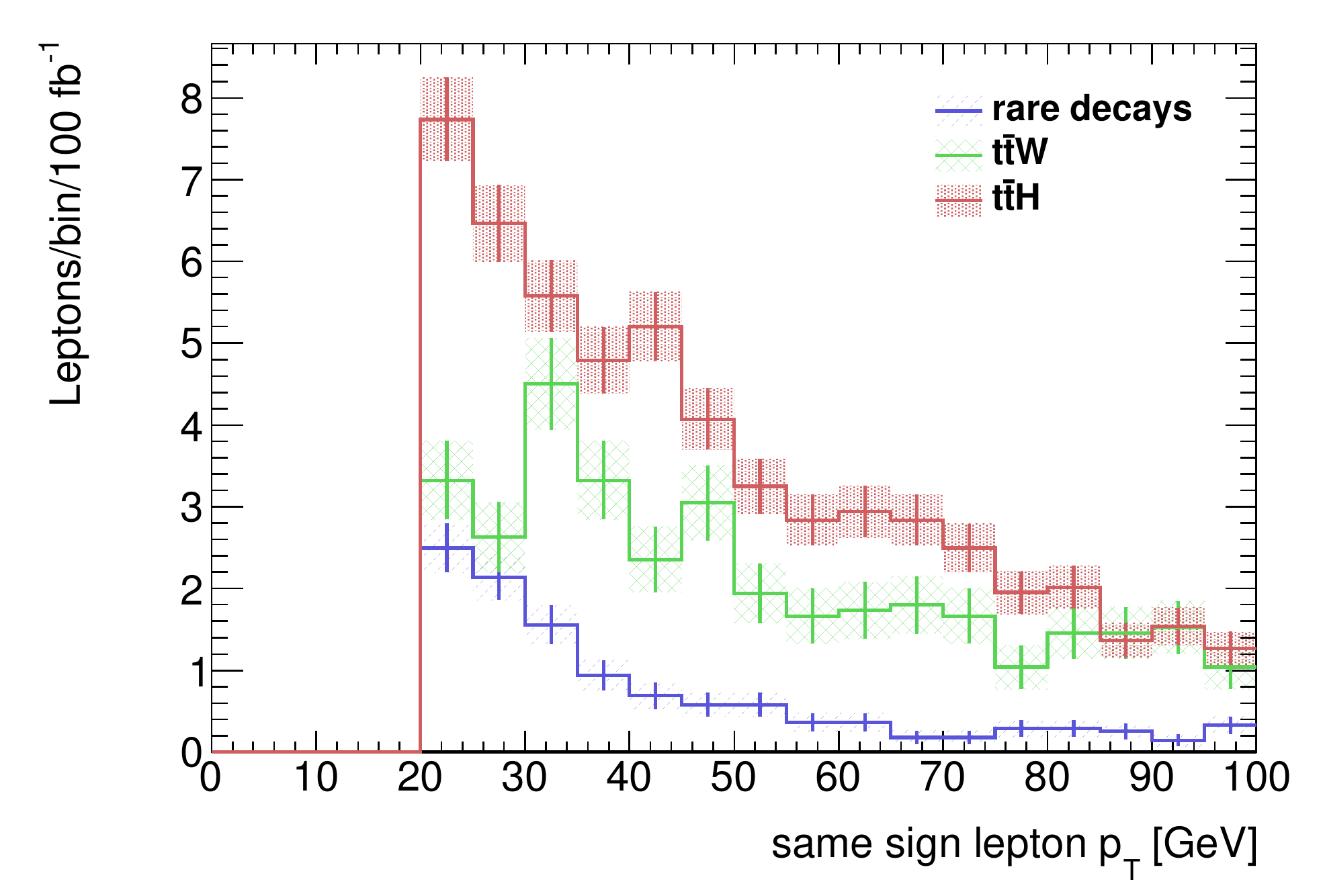}
  \caption{\label{fig:02sspt}}
 \end{subfigure}
 \begin{subfigure}[t]{.5\linewidth}
  \includegraphics[width=\linewidth]{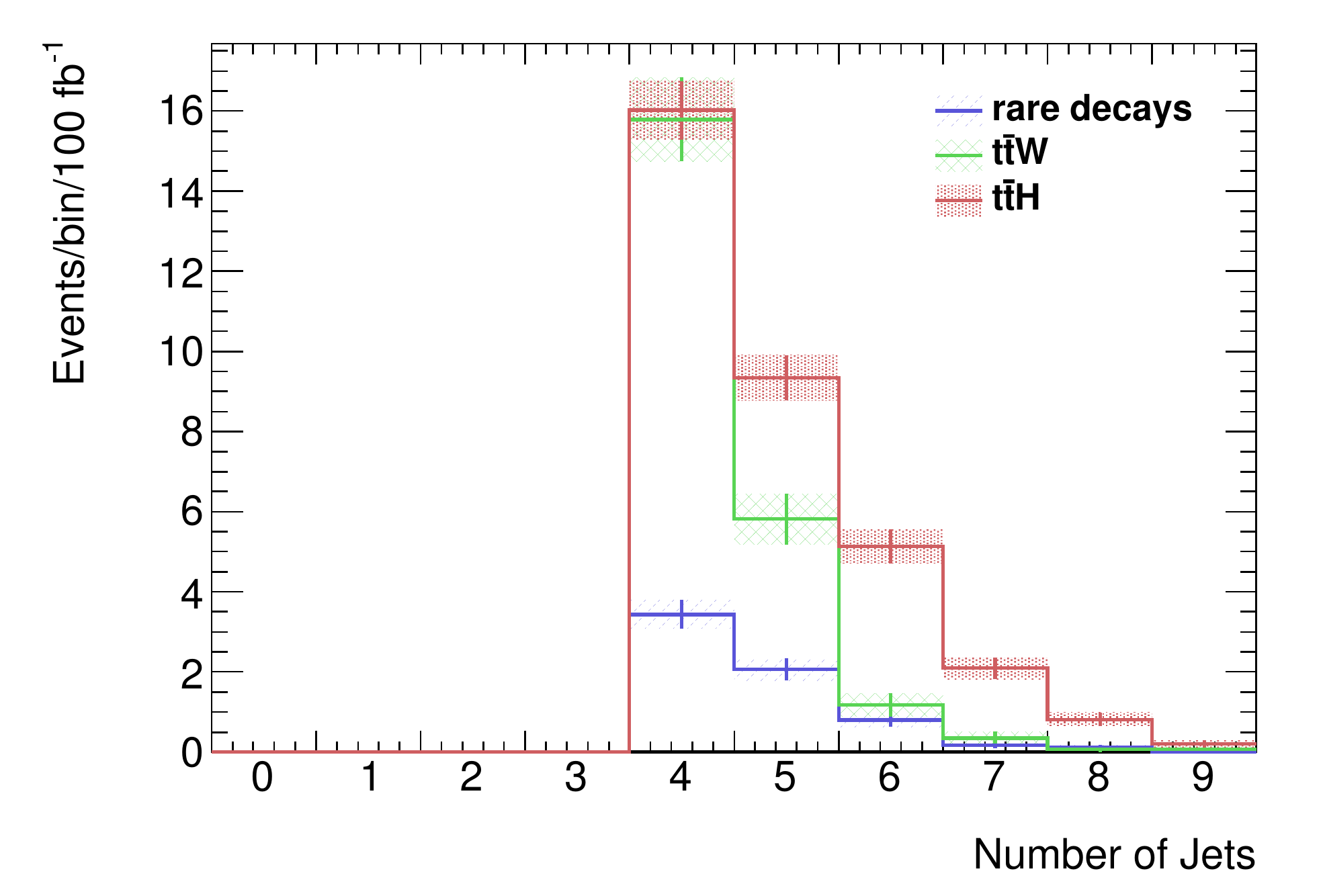}
  \caption{\label{fig:02njet}}
 \end{subfigure}
 \begin{subfigure}[t]{.5\linewidth}
  \includegraphics[width=\linewidth]{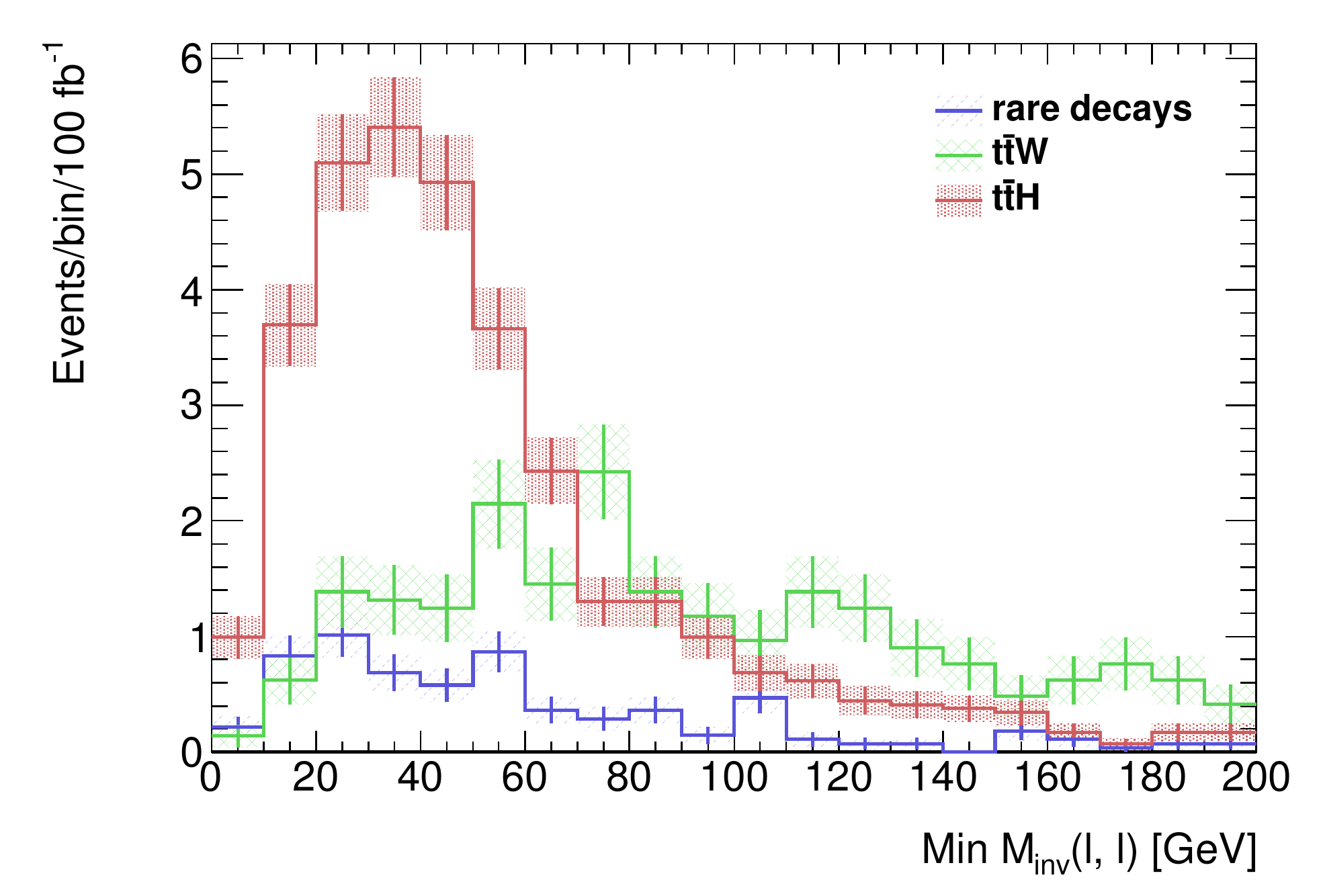}
  \caption{\label{fig:02mllmin}}
 \end{subfigure}
 \begin{subfigure}[t]{.5\linewidth}
  \includegraphics[width=\linewidth]{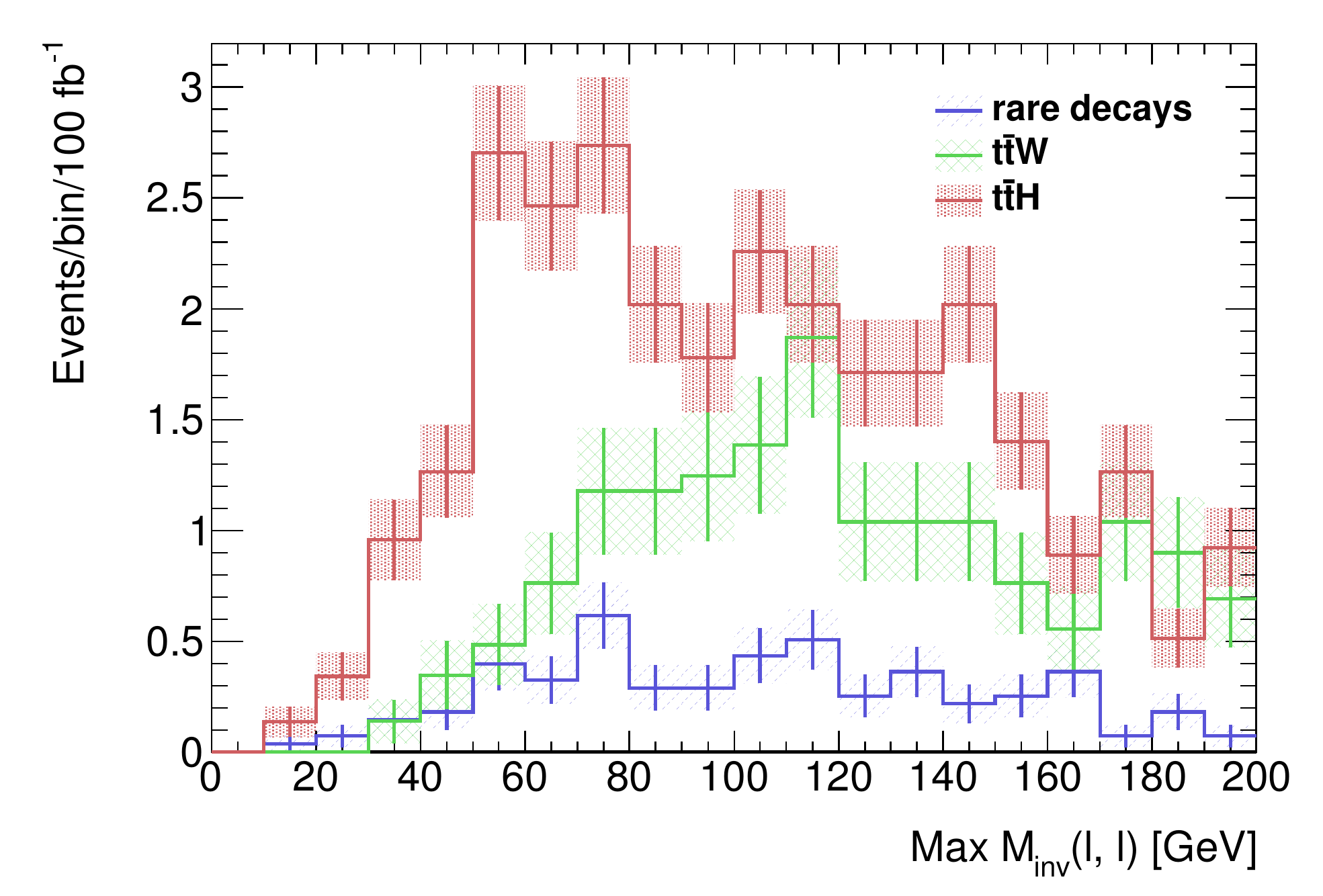}
  \caption{\label{fig:02mllmax}}
 \end{subfigure}
 \caption{\label{fig:scen3}Plots for various variables in the Scenario 3 selection: a) $p_T$ of the two selected leptons of the same charge; b) number of reconstructed jets with $p_T > 25$ GeV; c) the smaller of the two possible opposite sign dilepton masses $M(\ell^+\ell^-)$; d) the larger of the two possible opposite sign dilepton masses $M(\ell^+\ell^-)$. Yields are scaled to 100 fb$^{-1}$. Error bands correspond to statistical uncertainties in the simulated samples.}
\end{figure}

\begin{table}
\centering
\caption{\label{tbl:yields}Expected yields in 100 fb$^{-1}$ of 13 TeV data for rare $t\bar t$ decay with asymmetric internal conversion, $t\bar t W$, and $t\bar t H$ under three different trilepton selections, detailed in the text.  The rare decay contribution is normalized to a total production rate of 10.9 fb and has large uncertainties coming from higher-order corrections and the virtual photon invariant mass cutoff.}
 \begin{tabular}{lccc}
  \hline\hline
  Selection & $t\bar t$ rare decay & $t\bar t W$ & $t \bar t H$ \\
  \hline
  Scenario 1 & 106 & 244 & 91\\
  Scenario 2 & 26 & 170 & 54 \\
  Scenario 3 & 6.6 & 23 & 34\\
  \hline\hline
 \end{tabular}

\end{table}

Finally, we make a requirement on the total number of jets in order to isolate $t\bar t H$ and separate it from $t\bar t W$.  Some variables are shown in Figure~\ref{fig:scen3}.  Other than in the shape of the same-sign lepton $p_T$, where the same kink at 40 GeV is seen as in Scenario 2, it is hard to convincingly discriminate the rare decay and $t\bar t H$. 

The yields of the three considered processes under each scenario are shown in Table~\ref{tbl:yields}.  In Scenarios 2 and 3 ($t\bar t W$- and $t\bar t H$-search-like regions) the $t\bar t$ AIC background contributes 15--20\% of the relevant signal yield, and can have a non-trivial impact on extraction of $t\bar t W$ and $t\bar t H$ cross sections.  Due to the large uncertainty in the branching fraction for the rare decay, this impact may in fact be larger than seen here.

We raise one additional concern.  The rare decay process will also appear in
same-sign dilepton analyses, as the AIC lepton is not charge-correlated with
the leptons from $W$ decay in top events.  The AIC leptons will appear as a
relatively low $p_T$ contribution, and may be hard to distinguish from
non-prompt electrons and muons from hadronic decays that happen to be
isolated.  Extractions of fake factors and validation of data-driven
non-prompt estimates may be affected if this additional prompt lepton
contribution is not considered.  As non-prompt lepton rates are very
process- and event selection-specific, we do not discuss them further here.

\section{\label{sec:conclusion}Conclusion}
The rare top decays $t \to (\ell' \nu, q q') b \ell \ell$ have not generally
been considered as potential backgrounds for searches.  However, at low
$\ell\ell$ invariant mass, lepton kinematics and analysis selections can
result in a high likelihood of one of the two leptons being reconstructed; the
relatively high top decay branching fraction in this region may yield a
significant number of events in signal regions for $ttW$ or $ttH$
production. Considering the fact that there are persistent excesses in $t\bar
t W$ and $t\bar t H$ results over the Standard Model expectations in
multilepton searches, this background source should be carefully evaluated for
its impact in real LHC analyses.  A reliable calculation of the decay
branching fraction and kinematics, with a proper treatment of the IR
divergence as $m(\ell\ell) \to 0$, will be extremely valuable. Prescriptions
for matching QED parton shower calculations including $\gamma \to \ell\ell$
splitting and the corresponding perturbative matrix element calculations will
enable consistent simulations of these processes by the LHC experiments. A degree of data-driven background estimation is possible, requiring the use of soft leptons with momentum down to $\approx 5$ GeV to fully reconstruct symmetric internal conversion events.


\acknowledgments
We wish to thank Can K{\i}l{\i}\c{c} for invaluable discussion about
implementing the rare decays in \textsc{MG5\_aMC}. We also thank Tamara Vazquez
Schr\"oder and Josh McFayden for input and advice. This work was supported by
the US Department of Energy, Office of Science, Office of High Energy Physics,
under Award Number DE-SC0007890. This work used the Extreme Science and
Engineering Discovery Environment (XSEDE) \cite{XSEDE}, which is supported by US National Science Foundation grant number ACI-1548562. The authors acknowledge the Texas Advanced Computing Center (TACC) at the University of Texas at Austin for providing HPC resources that contributed to the results in this paper. 



\bibliographystyle{JHEP}
\bibliography{raretop}






\end{document}